\documentclass[%
reprint,
superscriptaddress,
amsmath,amssymb,
aps,
prb,
]{revtex4-2}

\usepackage{graphicx}% Include figure files
\usepackage[pdftex]{color} % need
\usepackage{dcolumn}% Align table columns on decimal point
\usepackage{bm}% bold math
\usepackage{here}
\usepackage[utf8]{inputenc}
\usepackage{color}
\usepackage[pdftex]{hyperref}% add hypertext capabilities 
\hypersetup{
colorlinks=true,
linkcolor=blue,
citecolor=blue,
urlcolor=blue,
setpagesize=false
}
\newcommand{\vk}{\bm{k}}
\newcommand{\vq}{\bm{q}}
\newcommand{\ave}[1]{\langle #1 \rangle}
\newcommand{\abs}[1]{\left | #1 \right |}
\usepackage{braket}

\begin{document}

%\preprint{}

\title{Piezoelectric effect and diode effect in anapole and monopole superconductors}
%\thanks{A footnote to the article title}

\author{Michiya Chazono}
\email[]{chazono.michiya.84s@st.kyoto-u.ac.jp}
\affiliation{Department of Physics, Kyoto University, Kyoto 606-8502, Japan}

\author{Shota Kanasugi}
%\email[]{...}
\affiliation{Department of Physics, Kyoto University, Kyoto 606-8502, Japan}

\author{Taisei Kitamura}
%\email[]{...}
\affiliation{Department of Physics, Kyoto University, Kyoto 606-8502, Japan}

\author{Youichi Yanase}
%\email[]{yanase@scphys.kyoto-u.ac.jp}
\affiliation{Department of Physics, Kyoto University, Kyoto 606-8502, Japan}
%\affiliation{Institute for Molecular Science, Okazaki 444-8585, Japan}

% \affiliation{......}

\date{\today}

%%%%%%%%% MAIN %%%%%%%%%

\begin{abstract}
Superconductors lacking both inversion symmetry and time-reversal symmetry have been attracting much attention as a platform for exotic superconducting phases and anomalous phenomena, including the superconducting diode effect. Recent studies revealed intrinsic phases with this symmetry, named anapole superconductivity and monopole superconductivity, which are $PT$-symmetric superconducting states with and without Cooper pairs' total momentum, respectively. %where the Bogoliubov spectrum can be asymmetric. 
To explore characteristic phenomena in these states, we calculate and predict the superconducting piezoelectric effect and superconducting diode effect. A close relationship with the finite-$q$ pairing, asymmetric Bogoliubov spectrum, and quantum geometry is discussed. This study reveals the piezoelectric and diode effects as potential probes to elucidate exotic superconducting phases.
\end{abstract}

\maketitle

%%%%%%%%%%%%%%%% Introduction %%%%%%%%%%%%%%%%
\section{INTRODUCTION}
In recent years, superconductors lacking both inversion symmetry (IS) and time-reversal symmetry (TRS) are received much attention. For instance, nonreciprocal charge responses are extensively studied in superconductors with such symmetry~\cite{Wakatsuki2017, Wakatsuki2018, Hoshino2018, Itahashi2020, Schumann2020, Nakamura2020, Miyasaka2021, Watanabe2022o, Watanabe2022m}. Especially, the superconducting diode effect has become a central topic in condensed matter physics, and vast experimental and theoretical studies are conducted~\cite{Ando2020, Lyu2021, Bauriedl2022, Baumgartner2022, Wu2022, Narita2022, Lin2021, Kopasov2021, Yuan2022, Daido2022, He2022, Ilic2022, Legg2022, Scammell2022, Jiang2022, Daido2022-2,Kawarazaki2022-pd}. These nonreciprocal phenomena are expected to be ubiquitous since the simultaneous breaking of IS and TRS can be realized in various situations. Superconductors with noncentrosymmetric crystal structures under magnetic fields are typical examples~\cite{Wakatsuki2017, Wakatsuki2018, Hoshino2018, Itahashi2020, Schumann2020, Ando2020, Lyu2021, Bauriedl2022, Yuan2022, Daido2022, He2022, Ilic2022, Legg2022, Daido2022-2,Kawarazaki2022-pd}. Spontaneous breaking of TRS symmetry due to magnetism also results in the superconducting diode effect~\cite{Narita2022, Lin2021, Scammell2022}.
%, Wysokinski2019}.
Furthermore, we can utilize the supercurrent that breaks both IS and TRS without dissipation~\cite{Nakamura2020, Miyasaka2021}. 

In the recent study~\cite{Ishizuka2021}, an intrinsic mechanism of spontaneous IS and TRS breaking is predicted for one of the multiple superconducting phases in UTe$_2$~\cite{Ran2019, Braithwaite2019, Aoki2020, Aoki2022, Rasuel2022, Kinjo2022, Sakai2022}. The competing instability of spin-triplet and spin-singlet superconductivity causes %\cite{Sundar2019, Knafo2021, Duan2021}, it is predicted that 
spontaneous parity mixing of Cooper pairs. %under an intermediate pressure. 
In the centrosymmetric crystals such as UTe$_2$, the phase difference between the even-parity and odd-parity pair potentials is likely to be $\pm \pi/2$ \cite{Wang2017}, leading to the IS and TRS breaking with intact $PT$-symmetry in the mixed-parity superconducting state. 
The $PT$-symmetric superconducting state is a novel quantum condensed phase of matter, and the realization in UTe$_2$ and other exotic superconductors is attracting attention. %expected to be the first superconductor with this mechanism. 
Therefore, it is eagerly desired to clarify the unique properties of the $PT$-symmetric superconducting states and to explore possible probes of them.
%superconducting phase diagram of UTe$_2$. 
%and verify the possible symmetry breaking. 
%to determine its unique phase diagram.

%\YYS{To identify the intrinsic IS and TRS breaking in the $PT$-symmetric superconducting state, the multiband effects may be essential. 
%In $PT$-symmetric multiband superconducting states with the above mechanism, 
%For example, it was shown that the Bogoliubov spectrum can be asymmetric due to the multiband effect \cite{Kanasugi2022, Kitamura2022}.} 

%\YYS{An ubiquitous feature of the $PT$-symmetric superconducting states is the asymmetric spectrum of Bogoliubov quasiparticles \cite{Kanasugi2022, Kitamura2022}, which arises from the multiband effects.}

Let us classify the $PT$-symmetric superconducting states. In analogy with the $PT$-symmetric magnetic order \cite{Watanabe2017}, they are classified into monopole, anapole, quadrupole, and higher-order multipole superconducting states. An intriguing class is the \textit{anapole superconductivity}, where Cooper pairs can get finite total momentum at zero magnetic field \cite{Kanasugi2022, Kitamura2022}. The anapole superconducting state is distinguished by the Fulde-Ferrell-Larkin-Ovchinnikov (FFLO) state \cite{FF1964, LO1964} and the helical superconducting state \cite{Bauer2012, Smidman2017} which require finite magnetic field or spin polarization. %From the viewpoint of symmetry, such finite-q paring state is called \textit{anapole superconducting state}, and it 
In the anapole superconductors, the finite-$q$ pairing state is characterized by a polar vector named the effective anapole moment, which was recently revealed to arise from various origins \cite{Kanasugi2022, Kitamura2022}. On the other hand, the other classes of $PT$-symmetric superconducting states are nonpolar, and Cooper pairs condensate with zero total momentum. %in a nonpolar system, 
%Such superconducting states are classified in analogy with the $PT$-symmetric magnetic order, and 
An example of them appears in the classification table for UTe$_2$ 
(Table~\ref{table:PT}),
named (\textit{magnetic}) \textit{monopole superconductivity}. 
The analysis of the periodic Anderson model has shown that both anapole and monopole superconducting states %are almost degenerate, and therefore, both of them https://ja.overleaf.com/project/62d60a5d96848b7305956432
are candidate superconducting states of UTe$_2$~\cite{Ishizuka2021}. 
%Therefore, it is desirable to verify the spontaneous IS and TRS breaking and furthermore identify the anapole or monopole superconductivity. 

%For these purposes, we would like to explore the characteristic phenomena which are allowed only when symmetry breaking occurs. 
An important consequence of the IS and TRS  breaking is the asymmetric spectrum of Bogoliubov quasiparticles, which arises from the multiband effects in anapole and monopole superconductors~\cite{Kanasugi2022, Kitamura2022}. In principle, we can distinguish all the $PT$-symmetric superconducting phases by the asymmetric profile of the Bogoliubov spectrum. However, the direct measurement of the Bogoliubov spectrum is challenging, especially for low-temperature superconductors. Thus, we are motivated to explore the macroscopic phenomena of anapole and monopole superconductors, especially those allowed only when symmetry breaking occurs. 

%Motivated by these researches, 
For this purpose, in this paper, 
we study the superconducting piezoelectric effect (SCPE), a supercurrent-induced lattice distortion that occurs only when the superconductors lack both IS and TRS. %in the anapole and monopole superconducting states. In our previous work, 
We have investigated the SCPE in two-dimensional helical superconductors \cite{Chazono2022}, where TRS is broken by an external magnetic field. This paper focuses on the SCPE in the anapole and monopole superconductors, where IS and TRS are broken by spontaneous parity mixing in Cooper pairs without noncentrosymmetric crystal structure or external field. We also investigate the superconducting diode effect (SDE) and predict the intrinsic SDE in the anapole superconducting state, although it is absent in the monopole superconducting state.
%which is also considered to be intrinsic in them. 
%We note that it is pointed out that both 
As a link of the SCPE and SDE with Cooper pairs' momentum was shown in the helical superconductors~\cite{Chazono2022, Yuan2022, Daido2022, He2022, Daido2022-2}, we also discuss the properties of finite-$q$ pairing in the anapole superconducting state for comparison. 
It is shown that the SCPE shows distinct behaviors depending on the origin of finite-$q$ pairing, namely, the asymmetric Bogoliubov spectrum and quantum geometry~\cite{Kanasugi2022,Kitamura2022}.
%As the finite-$q$ pairing is induced by the asymmetric Bogoliubov spectrum and quantum geometry~\cite{Kanasugi2022,Kitamura2022}, 

This paper is organized as follows. In Sec.~\hyperref[sec:Formulation]{II}, we show a minimal model for the anapole and monopole superconducting states introduced in the previous study~\cite{Kanasugi2022} and formulate the SCPE based on the model. We also introduce the classification of possible $PT$-symmetric superconducting states in UTe$_2$. %established in our recent work \cite{Kitamura2022}.  
In Sec.~\hyperref[sec:Result]{III}, we show numerical results of the SCPE coefficients and Cooper pairs' momentum. It is revealed that the SCPE occurs in both anapole and monopole superconducting states. 
The behaviors of the SCPE are closely related to the origin of the effective anapole moment, which causes finite-$q$ pairing.  
%regardless of the origin of the effective anapole moment. 
We demonstrate the SDE in Sec.~\hyperref[sec:SDE]{IV} as another anomalous phenomenon in the anapole superconductors. In Sec.~\hyperref[sec:Summary]{V}, we summarize our study and discuss the future outlook.
%%%%%%%%%%%%%%%% End of  Introduction %%%%%%%%%%%%%%%%

%%%%%%%%%%%%%%%% Formulation %%%%%%%%%%%%%%%%
\section{FORMULATION}
\label{sec:Formulation}
%%%%%%%%%%%% Minimal anapole model for UTe$_2$ %%%%%%%%%%%%%
\subsection{Minimal model for anapole and monopole superconducting states}
We adopt a minimal model for the anapole and monopole superconducting states, which was introduced in the previous study for UTe$_2$~\cite{Kanasugi2022}. While UTe$_2$ has a centrosymmetric crystal structure with $D_{2h}$ point group symmetry, the IS is locally broken on U sites owing to the sublattice degree of freedom. Using the Nambu spinor $\hat{c}^T_{\vk} = ( c_{\vk 1 \uparrow}, c_{\vk 2 \uparrow}, c_{\vk 1 \downarrow}, c_{\vk 2 \downarrow} )$ where $1, 2$ ($\uparrow, \downarrow$) denote the sublattice (spin) degree of freedom, we write the Bogoliubov-de Gennes (BdG) Hamiltonian in the following form:
%%%%BdGHamiltonian
\begin{align}  
\begin{footnotesize}
\mathcal{H}_{\vq} = 
\frac{1}{2} \sum_{\vk}
(
\begin{array}{cc}
\hat{c}^\dag_{\vk+\vq} & \hat{c}^T_{-\vk+\vq}
\end{array}
)
H_\textrm{BdG} (\vk,\vq)
\left(
\begin{array}{c}
\hat{c}_{\vk+\vq} \\
\hat{c}^*_{-\vk+\vq}
\end{array}
\right),
\label{BdGHamiltonian}
\end{footnotesize}
\\
%%%%BdGHamiltonianM
\begin{footnotesize}
H_\textrm{BdG} (\vk,\vq) =
\left(
\begin{array}{cc}
    H_0 (\vk+\vq) & \Delta (\vk) (i \sigma_y \otimes \tau_0) \\
   \left( \Delta (\vk) (i \sigma_y \otimes \tau_0) \right)^\dag  & - H_0 (-\vk+\vq)^T 
\end{array}
\right),
\label{BdGHamiltonianM}
\end{footnotesize}
\end{align}
where $\bm{\tau}$ ($\bm{\sigma}$) represents the Pauli matrix vector for the sublattice (spin) degree of freedom. We here assume Cooper pairs' total momentum $2 \vq$. Later, we show that $\vq$ is finite in the anapole superconducting state while $\vq = 0$ in the monopole superconducting state. The normal state Hamiltonian $H_0 (\vk)$ and superconducting order parameter $\Delta (\vk)$ are defined as follows. 

The sublattice degree of freedom with a locally noncentrosymmetric crystal structure allows the staggered antisymmetric spin-orbit coupling  (ASOC) in the centrosymmetric materials~\cite{LNSC_review}. Therefore, the normal state Hamiltonian is given by 
%%%%NHamiltonian
\begin{align}
H_0 (\vk) = (\varepsilon_{\vk} -\mu) \sigma_0 \otimes \tau_0 + \bm{g}_{\vk} \cdot \bm{\sigma} \otimes \tau_z, 
\label{NHamiltonian}
\end{align}
where $\varepsilon_{\vk} = - 2t (\cos{k_x} + \cos{k_y} + \cos{k_z})$ is a kinetic energy and $\bm{g}_{\vk} = (\alpha_x \sin{k_y},~ \alpha_y \sin{k_x},~ 0)$ represents the ASOC. Note that the relation $\alpha_y = - \alpha_x$ is not required because the local symmetry on U sites is orthorhombic $C_{2v}$, although it must be satisfied in the tetragonal $C_{4v}$ systems.

We consider mixed-parity order parameters for the anapole and monopole superconducting states, %even-parity Cooper pairing and odd-parity one simultaneously occur, 
and even- and odd-parity gap functions have $\pi/2$ phase difference consistent with the thermodynamic stability~\cite{Wang2017}. Because of the locally noncentrosymmetric crystal structure~\cite{LNSC_review}, in addition to the conventional even-parity spin-singlet and odd-parity spin-triplet pairings, even-parity spin-triplet and odd-parity spin-singlet pairings can be finite. To satisfy the fermion's anti-commutation relation, the superconducting order parameter is given by
%%%%SCHamiltonian
\begin{align}
\Delta (\vk) &=  \Delta_1 (\psi^g_{\vk} \sigma_0 \otimes \tau_0 + \beta \bm{d}^g_{\vk} \cdot \bm{\sigma} \otimes \tau_z) \nonumber \\
&+ \Delta_2 (\psi^u_{\vk} \sigma_0 \otimes \tau_z + \bm{d}^u_{\vk} \cdot \bm{\sigma} \otimes \tau_0) ,
\label{SCHamiltonian}
\end{align}
where $\psi^{g(u)}_{\vk}$ is an even-parity (odd-parity) spin-singlet component and  $\bm{d}^{g(u)}_{\vk}$ is an even-parity (odd-parity) spin-triplet component of the gap function. 
In this paper, 
%(an odd-parity spin-singlet component is ignored for simplicity). 
%Based on the previous study \cite{Ishizuka2021}, 
we assume the even-parity component belonging to the $A_g$ irreducible representation, specifically, $\psi^g_{\vk} = 1$ and $\bm{d}^g_{\vk} = (0, \sin{k_x}, 0)$. For the odd-parity component,  $B_{3u}$ and $A_u$ irreducible representations are examined.  For simplicity, $\psi^u_{\vk}=0$ is ignored, and we consider spin-triplet pairing with $\bm{d}^u_{\vk} = (0, 0, \sin{k_y})$ and $\bm{d}^u_{\vk} = (0, 0, \sin{k_z})$, corresponding to the $B_{3u}$ and $A_u$ representations, respectively. 
We choose a real $\Delta_1$ and a pure imaginary $\Delta_2$ without loss of generality. %to consider the $PT$-symmetric superconducting state.
%The point group of the coexistent $A_g + i B_{3u}$ state is polar $C_{2v}$, while that of the $A_g + i A_u$ state is nonpolar $D_2$. Thus, the former is the anapole superconducting state and the latter is the monopole superconducting state in analogy with the classification of magnetic states~\cite{Watanabe2018}. We show the classification of $PT$-symmetric superconducting states in Table \ref{table:AM}. 
The coexistent $A_g + i B_{3u}$ state and $A_g + i A_u$ state realize the anapole and monopole superconductivity, as discussed in the next subsection. 
%Our choice of the representations is in accordance with the 
These states are predicted in a microscopic analysis of the periodic Andeson model for UTe$_2$ \cite{Ishizuka2021}. 
%where the $A_g + i B_{3u}$ and $A_g + i A_u$ states are predicted. 
However, the following results are expected to be general in the sense that the main conclusion for the SCPE and SDE applies to other representations, such as the $A_g + i B_{1u}$ state, as well. 
For a later discussion, we introduce $\beta$ as the strength of 
the staggered spin-triplet gap function allowed in locally noncentrosymmetric superconductors \cite{LNSC_review}. 
We will see that $\beta$ is an important parameter for the superconducting properties. 
%\YYS{We note that there are no inter-sublattice terms ($\tau_x$ and $\tau_y$) in our model. Therefore, parameter sets ($\alpha_x, \alpha_y, \beta$) and ($-\alpha_x, -\alpha_y, -\beta$) give the same results while the difference of their relative sign can be essential. (This part is moved to later.)}
%A$\bm{d}^g_{\vk}$. 
%%%%%%%%%%%% End of Minimal anapole model for UTe$_2$ %%%%%%%%%%%%%

%%%%%%%%%%%% Superconducting piezoelectric effect (SCPE) %%%%%%%%%%%%%
\subsection{Superconducting piezoelectric effect (SCPE)}
The SCPE is defined as a lattice distortion induced by a supercurrent \cite{Chazono2022}. In the linear response regime, it is formulated as follows,
%%%%SCPE
\begin{align}
s_{ij} = d^{\textrm{(SC)}}_{ijk} J_k,
\label{SCPE}
\end{align}
where $s_{ij}$ is a strain tensor, $d^{\textrm{(SC)}}_{ijk}$ is a SCPE coefficient, and $J_k$ is a supercurrent. Because $s_{ij}$ ($J_k$) is a parity even (odd) and time-reversal even (odd) quantity, $d^{\textrm{(SC)}}_{ijk}$ can be finite only in systems lacking both IS and TRS. 

The structure of the SCPE tensor $d^{\textrm{(SC)}}_{ijk}$ depends on the point group of the superconducting states. %system what modes can be realized. 
For a coefficient $d^{\textrm{(SC)}}_{ijk}$ to be finite, corresponding $s_{ij}$ and $J_k$ must belong to the same irreducible representation. Let us consider the cases of our model. 
The point group of the $A_g + i B_{3u}$ state is polar $C_{2v}$, while that of the $A_g + i A_u$ state is nonpolar $D_2$. Thus, the former is the anapole superconducting state and the latter is the monopole superconducting state in analogy with the classification of magnetic states~\cite{Watanabe2018}. 
%anapole superconducting state and $A_g + i A_u$ monopole superconducting state, respectively.
%In this study, we consider $C_{2v}$ and $D_2$ systems which correspond to the $A_g + i B_{3u}$ anapole superconducting state and $A_g + i A_u$ monopole superconducting state, respectively. 
The classification of the strain $s_{ij}$ and supercurrent $J_k$ based on the $C_{2v}$ and $D_2$ point group is summarized in 
%The relationship is summarized in 
Table~\ref{table:Ana} and \ref{table:Mono}, respectively. 
The SCPE modes allowed by symmetry are also shown in the tables.
We show the classification of $PT$-symmetric superconducting states in Table \ref{table:PT} and see that all the $PT$-symmetric superconducting states belong to either $C_{2v}$ or $D_2$ point group in the orthorhombic $D_{2h}$ system. 
Thus, the classification in Table~\ref{table:Ana} and \ref{table:Mono} applies the other states as well, when we choose an appropriate two-fold rotation axis for $C_{2v}$ (See Appendix A for the complete results).
%We note that there exist other anapole and monopole states possible in $D_{2h}$ systems. We summarize them in Appendix A.

\begin{figure*}[htbp]
  \centering
  \includegraphics[width=160mm]{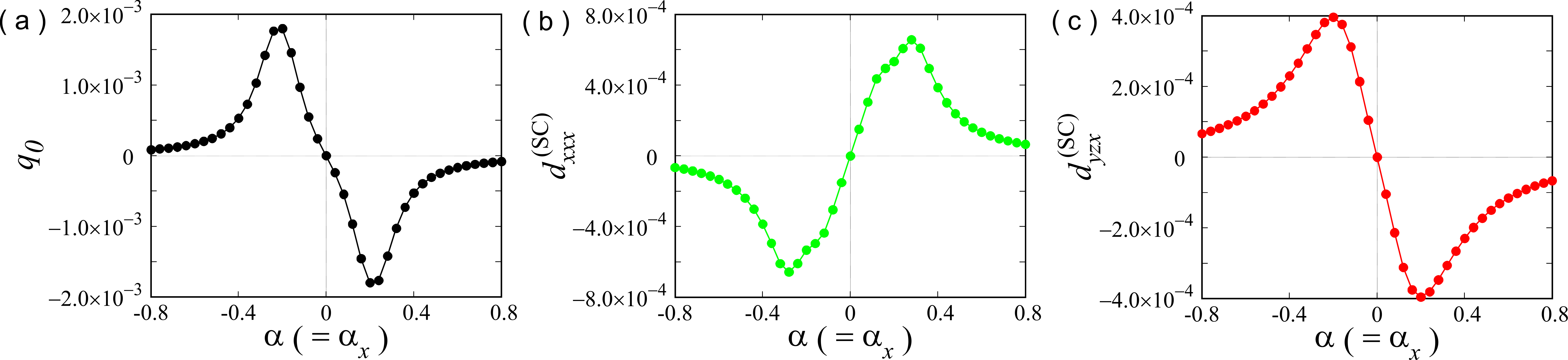}
  \caption{ 
 Results of the group velocity model: $\alpha (=\alpha_x)$ dependence of (a) $q_0$, (b) $d^{\textrm{(SC)}}_{xxx}$ in the $A_g + i B_{3u}$ anapole state, and (c) $d^{\textrm{(SC)}}_{yzx}$ in the $A_g + i A_{u}$ monopole state. We set $\alpha_y = 0,~ \beta = 1$ and $T = 0.01$.
 }
\label{fig:alp_ay0}
\end{figure*}

%%%%bet_ay0
\begin{figure*}[htbp]
  \centering
  \includegraphics[width=160mm]{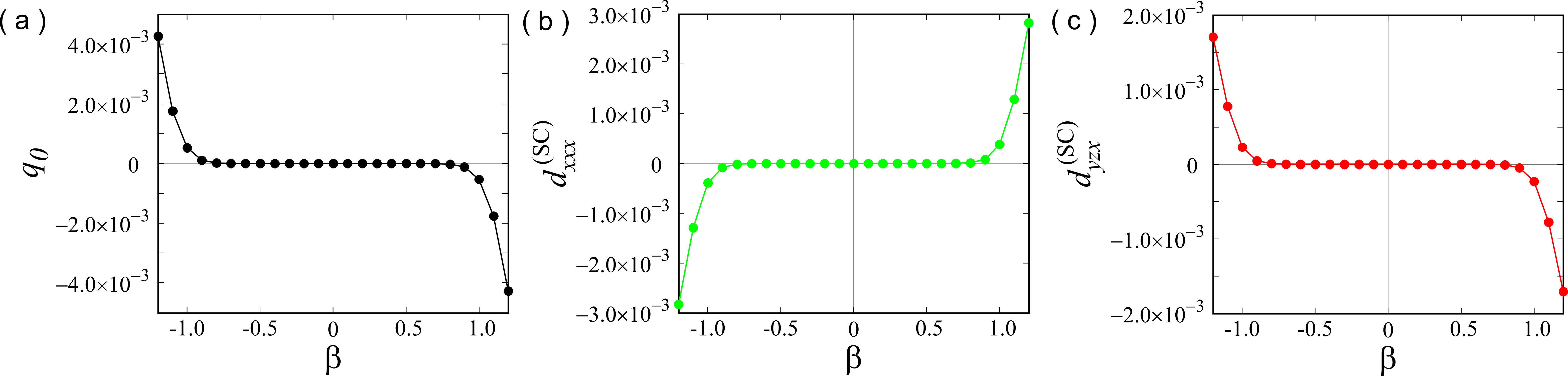}
  \caption{ 
 Results of the group velocity model: $\beta$ dependence of (a) $q_0$, (b) $d^{\textrm{(SC)}}_{xxx}$ in the $A_g + i B_{3u}$ anapole state, and (c) $d^{\textrm{(SC)}}_{yzx}$ in the $A_g + i A_{u}$ monopole state. We set $\alpha_y = 0,~ \alpha (=\alpha_x)= 0.4$ and $T = 0.01$.
 }
\label{fig:bet_ay0}
\end{figure*}

\begin{figure}[htbp]
  \centering
  \includegraphics[width=80mm]{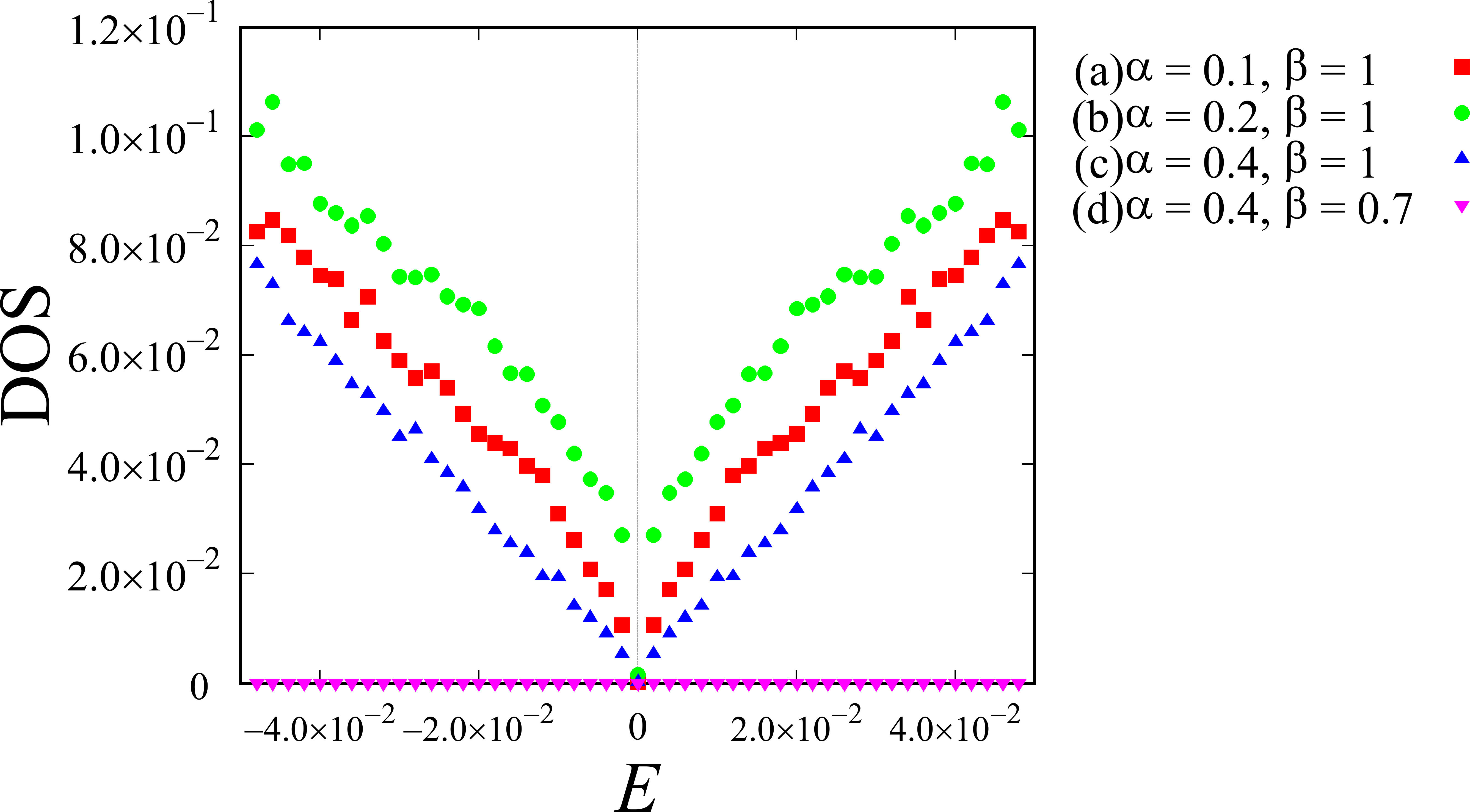}
  \caption{ The DOS in the group velocity model for the $A_g + i B_{3u}$ anapole state.
  We set $q_0=0$ %in this calculation 
  for simplicity.
  The parameters are (a) $\alpha = 0.1, \beta = 1$, (b) $\alpha = 0.2, \beta = 1$, (c) $\alpha = 0.4, \beta = 1$, and (d) $\alpha = 0.4, \beta = 0.7$. 
 }
\label{fig:enedos_ay0}
\end{figure}

%%%%table:Ana Mono
\renewcommand\arraystretch{1.5}
\begin{table}[t]
 \caption{List of the irreducible representations (IRs) of the $C_{2v}$ point group ($A_g + i B_{3u}$ anapole superconducting state) and corresponding strain $s_{ij}$, supercurrent $J_k$, and SCPE mode $d^{\textrm{(SC)}}_{ijk}$.}
 \label{table:Ana}
 \centering
  \begin{tabular}{ccc||c}
  \hline
   IR & Strain & Supercurrent & SCPE mode \\
  \hline \hline
   $A_{1}$ & $s_{xx}, s_{yy}, s_{zz}$ & $J_x$ & $d^{\textrm{(SC)}}_{xxx}~ d^{\textrm{(SC)}}_{yyx}~ d^{\textrm{(SC)}}_{zzx}$ \\
  \hline
   $A_{2}$ & $s_{yz}$ & $-$ & $-$ \\
  \hline
   $B_{1}$ & $s_{zx}$ & $J_z$ & $d^{\textrm{(SC)}}_{zxz}$ \\
  \hline
   $B_{2}$ & $s_{xy}$ & $J_y$ & $d^{\textrm{(SC)}}_{xyy}$ \\
  \hline
  \end{tabular}
\end{table}
\begin{table}[t]
 \caption{List of the IRs of the $D_{2}$ point group ($A_g + i A_{u}$ monopole superconducting state) and corresponding strain $s_{ij}$, supercurrent $J_k$, and SCPE mode $d^{\textrm{(SC)}}_{ijk}$.}
 \label{table:Mono}
 \centering
  \begin{tabular}{ccc||c}
  \hline
   IR & Strain & Supercurrent & SCPE mode \\
  \hline \hline
   $A_{1}$ & $s_{xx}, s_{yy}, s_{zz}$ & $-$ & $-$ \\
  \hline
   $B_{1}$ & $s_{xy}$ & $J_z$ & $d^{\textrm{(SC)}}_{xyz}$ \\
  \hline
   $B_{2}$ & $s_{zx}$ & $J_y$ & $d^{\textrm{(SC)}}_{zxy}$ \\
  \hline
   $B_{3}$ & $s_{yz}$ & $J_x$ & $d^{\textrm{(SC)}}_{yzx}$ \\
  \hline
  \end{tabular}
\end{table}

\begin{table}[htbp]
 \caption{Point group of the mixed-parity superconducting states in $D_{2h}$ systems. For the $C_{2v}$ point group, $x, y$ and $z$ denote the principal axis.}
 \label{table:PT}
 \centering
  \begin{tabular}{c||c|c|c|c}
  \hline
    & $A_g$ & $B_{1g}$ & $B_{2g}$ & $B_{3g}$ \\
  \hline \hline
   $A_u$ & $D_2$ & $C_{2v} (z)$ & $C_{2v} (y)$ & $C_{2v} (x)$ \\
  \hline
   $B_{1u}$ & $C_{2v} (z)$ & $D_2$ & $C_{2v} (x)$ & $C_{2v} (y)$ \\
  \hline
   $B_{2u}$ & $C_{2v} (y)$ & $C_{2v} (x)$ & $D_2$ & $C_{2v} (z)$\\
  \hline
   $B_{3u}$ & $C_{2v} (x)$ & $C_{2v} (y)$ & $C_{2v} (z)$ & $D_2$ \\
  \hline
  \end{tabular}
\end{table}

For the estimation of the strain, we calculate the expectation values of the weighted density operator, which characterizes the modulation of the hopping parameters due to the distortion~\cite{Watanabe2017, Chazono2022}. 
Although the strain is proportional to the weighted density, we avoid calculating the proportionality coefficient because it strongly depends on material parameters, such as electron-phonon coupling and elastic modulus.
%It is hard to calculate the actual strain $s_{ij}$ because it strongly depends on the material parameters, such as the electron-phonon coupling and elastic modulus. We here simplify the calculation using the weighted density operator which characterizes the modulation of the hopping parameters induced by the distortion \cite{Watanabe2017, Chazono2022}
%%%%wdo
The weighted density operator is defined as
\begin{align}
\hat{n}_{ij} = \sum_{\vk} 
(
\begin{array}{cc}
\hat{c}^\dag_{\vk+\vq} & \hat{c}^T_{-\vk+\vq}
\end{array}
)
n_{ij} (\vk,\vq)
\left(
\begin{array}{c}
\hat{c}_{\vk+\vq} \\
\hat{c}^*_{-\vk+\vq}
\end{array}
\right),
\label{wdo}
\end{align}
%%%%nij
\begin{align}
\begin{footnotesize}
n_{ij} (\vk,\vq) = \frac{1}{2}
\left(
\begin{array}{cc}
    D_{ij} (\vk+\vq) \mathalpha{\times} I_4 & 0 \\
    0 & - D_{ij} (-\vk+\vq) \mathalpha{\times} I_4
\end{array}
\right),
\end{footnotesize}
\label{nij}
\end{align}
%%%%Dij
\begin{align}
D_{ij} (\vk)  = 
\begin{cases}
\cos{k_i} & (i = j) \\
\sin{k_i} \sin{k_j} & (i \ne j)
\end{cases},
\label{Dij}
\end{align}
%%%%nave
%We approximate $s_{ij}$ by the expectation value of $\hat{n}_{ij}$
and expectation values are calculated by
\begin{align}
&\ave{\hat{n}_{ij}}_{\textrm{eq}, \vq} \notag\\&= \frac{1}{V} \sum_{\vk \alpha} 
[U^\dag (\vk,\vq) n_{ij}(\vk,\vq) U (\vk,\vq)]_{\alpha \alpha} f ( E_\alpha (\vk,\vq) ),
\label{nave}
\end{align}
where $E_\alpha$ are eigenvalues of the Hamiltonian, Eq.~\eqref{BdGHamiltonianM}
%%%%Ealpha
\begin{align}
E_\alpha (\vk,\vq) = [U^\dag (\vk,\vq) H_\textrm{BdG} (\vk,\vq) U (\vk,\vq)]_{\alpha \alpha},
\label{Ealpha}
\end{align}
and $f (E) = ( e^{E/T} + 1 )^{-1}$ is the Fermi distribution function. We calculate the expectation value of the supercurrent in a similar way,
%%%%Jave
\begin{align}
&\ave{\hat{J}_{k}}_{\textrm{eq}, \vq} \notag\\&=\frac{1}{V} \sum_{\vk \alpha} 
[U^\dag (\vk,\vq) J_{k}(\vk,\vq) U (\vk,\vq)]_{\alpha \alpha} f ( E_\alpha (\vk,\vq) ),
\label{Jave}
\end{align}
where
%%%%Jk
\begin{align}
\begin{small}
J_{k} (\vk,\vq) = \frac{e}{2}
\left(
\begin{array}{cc}
    \partial_{k_k} H_\textrm{0}(\vk+\vq) & 0 \\
    0 & - \partial_{k_k} H_\textrm{0}(-\vk+\vq)
\end{array}
\right).
\end{small}
\label{Jk}
\end{align}
Then, we redefine the SCPE coefficient $d^{\textrm{(SC)}}_{ijk}$ by
%%%%dijk
\begin{align}
d^\textrm{(SC)}_{ijk} = \lim_{q^\prime_k \to 0}
\frac{\ave{\hat{n}_{ij}}_{\text{eq}, \vq + q^\prime_k} - \ave{\hat{n}_{ij}}_{\text{eq}, \vq - q^\prime_k}}{\ave{\hat{J}_k}_{\text{eq}, \vq + q^\prime_k} - \ave{\hat{J}_k}_{\text{eq}, \vq - q^\prime_k}},
\label{dijk}
\end{align}
where $q^\prime_k$ is Cooper pairs' momentum parallel to the supercurrent $J_k$.
%%%%%%%%%%%% End of Superconducting piezoelectric effect (SCPE) %%%%%%%%%%%%%

%%%%%%%%%%%% Classification of the effective anapole moment %%%%%%%%%%%%%
\subsection{Classification of anapole superconducting states}\label{subsec:classification_anapole}

In the previous section, the superconducting states have been classified based on symmetry. Here, we furthermore classify the anapole superconducting states by their microscopic properties.

The effective anapole moment in the superconducting state is defined by the first derivative of thermodynamic potential with respect to the Cooper pairs' momentum \cite{Kanasugi2022}, and thus, finite anapole moment directly indicates the finite-$q$ pairing state.
As clarified in our recent work, there exist several origins of the effective anapole moment \cite{Kitamura2022}. They are classified into the group velocity term and the geometric term. While the former arises from the asymmetric Bogoliubov spectrum as pointed out in the previous study \cite{Kanasugi2022}, the latter is induced by the quantum geometric effect, which recently attracts attention in various fields~\cite{marzari1997maximally,resta2011the,peotta2015superfluidity,liang2017band,torma2022superconductivity,gao2014field,gao2019nonreciprocal,lapa2019semiclassical,daido2020thermodynamic,julku2021excitations,julku2021quantum,ahn2020low,watanabe2021chiral,rhim2020quantum,solnyshkov2021quantum,liao2021experimental}. 
Owing to the geometric term, the finite-$q$ pairing state can be stabilized even for ordinary electronic states, where neither the  asymmetry nor the Zeeman splitting exists.
%量子幾何の説明が必要かと思ったので，代表的な文献だけ追加しておきました．この説明が必要かどうかは茶園さんの方で判断していいただければ助かります．
Thus, the anapole superconducting states can be classified into three cases: The anapole moment is owing to (1) purely group velocity term, (2) purely geometric term, and (3) cooperation of the two terms.

%Applying the classification to our model, it is classified as follows depending on the parameters:
The three cases can be represented in our model by choosing the following parameters:

(1) $\alpha_y = 0$ and $\beta \ne 0$ (Group velocity model),

(2) $\alpha_y \ne 0$ and $\beta = 0$ (Geometric effect model),

(3) $\alpha_y \ne 0$ and $\beta \ne 0$ (Mixed model).\\
In the group velocity model, the group velocity term is finite while the geometric term vanishes. %there is no group velocity term in the geometric effect model. 
In contrast, the group velocity term vanishes in the geometric effect model. Both the group velocity and geometric terms are finite in the mixed model. %We here assume the other parameters are not zero.
In the next section, we show that the behaviors of the SCPE are different between the three cases.

In this study, we adopt the following parameters unless we explicitly state otherwise: $t = 1$, $\mu = -4$, $\Delta_1 (T) = 0.2 \sqrt{1 - (T/T_{\rm c})}$, $\Delta_2 (T) = 0.2 i \sqrt{1 - (T/T_{\rm c})}$, (namely, $\abs{\Delta_2} = \Delta_1$), and the transition temperature is $T_{\rm c} = 0.1$.
%%%%%%%%%%%% End of Classification of the origin of anapole SC %%%%%%%%%%%%%

%%%%%%%%%%%%%%%% Result %%%%%%%%%%%%%%%%

%%%%tem_ay0
\begin{figure*}[htbp]
  \centering
  \includegraphics[width=160mm]{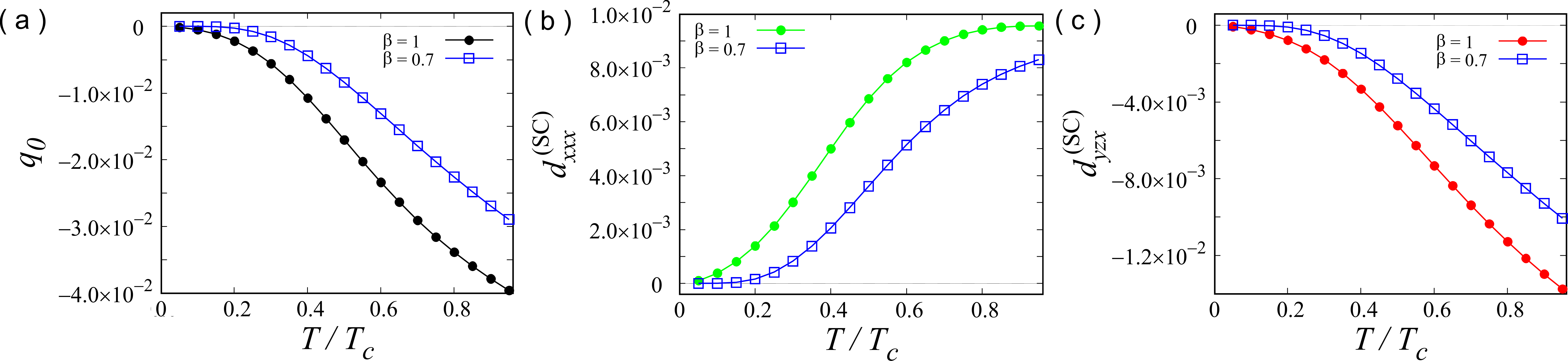}
  \caption{ 
 Results of the group velocity model: temperature dependence of (a) $q_0$, (b) $d^{\textrm{(SC)}}_{xxx}$ in the $A_g + i B_{3u}$ anapole state, and (c) $d^{\textrm{(SC)}}_{yzx}$ in the $A_g + i A_{u}$ monopole state. We set $\alpha_y = 0$, $\alpha (=\alpha_x) = 0.4$ and $\beta = 1$ or $0.7$.
 }
\label{fig:tem_ay0}
\end{figure*}

\section{RESULT : SCPE}
\label{sec:Result}
In this section, we show the numerical results of the Cooper pairs' momentum $\vq = (q_0, 0, 0)$, the SCPE coefficient  $d^{\textrm{(SC)}}_{xxx}$ in the $A_g + i B_{3u}$ anapole state, and $d^{\textrm{(SC)}}_{yzx}$ in the $A_g + i A_u$ monopole state. Because we find $\vq = 0$ in the monopole state, we show $q_0$ only for the anapole state.
We discuss the $\alpha (= \alpha_x)$, $\beta$, and $T$ dependence of the SCPE in the three models introduced in the previous section (Sec.~\ref{subsec:classification_anapole}) and compare them with the Cooper pairs' momentum in the anapole superconducting state.
A close relation between the SCPE and Cooper pairs' momentum is revealed.

%%%%%%%% group velocity model %%%%%%%%
\subsection*{(1) Group velocity model}

First, we analyze the group velocity model, where we set $\alpha_y = 0$ and $\beta \ne 0$.
%where the effective anapole moment originates from only the group velocity term. 
In this model, the effect of the quantum geometry of Bloch states is negligible. The $\alpha$ and $\beta$ dependences of $d^{\textrm{(SC)}}_{ijk}$ are shown in Figs.~\ref{fig:alp_ay0} and \ref{fig:bet_ay0}, together with the Cooper pairs' momentum $q_0$ in the anapole state. 
It is shown that the SCPE coefficients are finite in the anapole and monopole superconducting states and their magnitudes are comparable. 
Thus, we see the SCPE in the $PT$-symmetric superconducting states, irrespective of whether the Cooper pairs' momentum is finite or zero. 
%These results clarify that the existence of Cooper pairs' total momentum is not a necessary condition for the SCPE and we can expect the sizable SCPE even without it.  

%%%enedos_ay0

On the other hand, we notice similarities between the SCPE and Cooper pairs' momentum by comparing the parameter dependence of 
$d^{\textrm{(SC)}}_{ijk}$ and $q_0$. In particular, these quantities are antisymmetric with respect to $\alpha$ and $\beta$. The antisymmetric behavior of $q_0$ is expected from the result of the anapole moment $T_x$~\cite{Kitamura2022} because the relation $q_0 \approx - T_x / D_{xx}$ holds with $D_{xx}$ being the superfluid weight  and $T_x$ is $\alpha$ ($\beta$) antisymmetric in this model. On the other hand, it is nontrivial that $d^{\textrm{(SC)}}_{ijk}$ is also antisymmetric.
%Referring to our recent work \cite{Kitamura2022}, we can confirm that the anapole moment in this model is also $\alpha$- ($\beta$-) antisymmetric. Given that the anapole moment is the first derivative of free energy with respect to $\vq$ and concerned with supercurrent, it is natural that $d^{\textrm{(SC)}}_{ijk}$ and $q_0$ reflect its symmetry in the anapole state. 
We can interpret the similarities by analogy with the magnetopiezoelectric effect~\cite{Watanabe2017, Shiomi2019, Shiomi2020}, the counterpart of the SCPE in the normal state. It was shown that the magnetopiezoelectric effect originates from the asymmetric Fermi surface~\cite{Watanabe2017}.
In the group velocity model, the Cooper pairs' momentum arises from the asymmetric spectrum. 
Therefore, it is reasonable that the SCPE and Cooper pairs' momentum show similar behaviors.
%In the group velocity model, 
Indeed, their antisymmetric behaviors with respect to $\alpha$ and $\beta$ are explained as follows. The Bogoliubov spectrum is asymmetric and it is reversed by changing the sign of $\alpha$ or $\beta$ in this model (see Appendix B). Since the SCPE and Cooper pairs' momentum are caused by the asymmetric spectrum, it is natural that reversing spectrum changes the  sign of $d^{\textrm{(SC)}}_{ijk}$ and $q_0$. We stress that this interpretation is valid even for the monopole state.  

Furthermore, $d^{\textrm{(SC)}}_{ijk}$ and $q_0$ show a similar peak structure around $\alpha = \pm 0.2$ and drastically change around $\beta = \pm 1$. These behaviors are related to the density of states (DOS) in the low-energy region, which is shown for the $A_g + i B_{3u}$ anapole state with several parameters (Fig.~\ref{fig:enedos_ay0}). 
Note that the DOS in  Fig.~\ref{fig:enedos_ay0} is calculated with $q_0 =0$ for simplicity. Parameters leading to larger DOS around the Fermi level $E = 0$ correspond to larger $q_0$ and $d^{\textrm{(SC)}}_{ijk}$. This is consistent with the fact that the asymmetric spectrum is the main source of both $q_0$ and $d^{\textrm{(SC)}}_{ijk}$. It is also indicated that thermally excited quasiparticles are essential.
In Fig.~\ref{fig:enedos_ay0}, we see that the low-energy Bogoliubov spectrum is sensitive to the parameter $\beta$. With our parameter set, the spectrum is fully-gapped for $\abs{\beta} < 1$ while the nodal spectrum appears with $\abs{\beta} \ge 1$. To be more precise, point nodes are present at $\beta=1$, and when $\beta >1$, the Bogoliubov Fermi surface appears (see Appendix B), which has been studied with interest in the topological nature~\cite{Agterberg2017, Brydon2018, Link2020}. 
Thus, the low-energy DOS increases for $\beta \geq 1$, and therefore, the SCPE and Cooper pairs' momentum are enhanced.
%This property is consistent with the $\beta$ dependence shown in Fig~\ref{fig:bet_ay0}.}
%\YYS{We emphasize that there are accidental point nodes when $\abs{\beta} \ge 1$. The nodes and their surroundings turn into the Bogoliubov Fermi surface \cite{Agterberg2017, Brydon2018, Link2020} when we take into accout finite $q_0$. See Appendix B for more details.}
%%%%ABS_ay0
%\begin{figure}[tb]
%  \centering
%  \includegraphics[width=60mm]{ABS_ay0_woq.pdf}
%  \caption{ 
%The Bogoliubov spectrum of the $A_g + i B_{3u}$ state with following parameters : (a) $\alpha = 0.20, \beta = 1.0$, (b) $\alpha = 0.40, \beta = 1.0$, (c) $\alpha = 0.80, \beta = 1.0$ and  (d) $\alpha = 0.40, \beta = 0.70$. To emphasize the asymmetric structure, we draw the spectrum setting $k_z = 0$ and  $k_y = 0, \pi/8, \pi/4, 3\pi/8$.
% }
%\label{fig:ABS_ay0}
%\end{figure} 

Based on the above discussion, we expect a notable temperature dependence reflecting the gap structure.
Indeed, a characteristic behavior is observed in the temperature dependence of $q_0$ and $d^{\textrm{(SC)}}_{ijk}$, which are suppressed in the low-temperature region (Fig.~\ref{fig:tem_ay0}). 
%The suppression is a characteristic behavior of the group velocity model.
In Ref.~\cite{Kitamura2022}, it is shown that the asymmetry of the Bogoliubov spectrum is reflected in the anapole moment through the Fermi distribution function as $f(E(\vk)) - f(E(-\vk))$, %in a certain approximation.
and therefore, the anapole moment and $q_0$ are suppressed in the low-temperature region of gapped states.  %become small since $f(E(\vk)) \approx f(E(-\vk))$. 
The SCPE is expected to be suppressed by the same mechanism.
%This behavior suggests that a similar interpretation is valid for the SCPE.  
%
Indeed, the SCPE shows exponential temperature dependence in the full-gapped state ($\beta=0.7$) while it shows power-law dependence in the nodal state ($\beta=1$).
These results support the fact that the SCPE in the group velocity model %is closely concerned with the structure of the 
relies on the asymmetric energy spectrum, like the magnetopiezoelectric effect in the odd-parity magnetic ordered states~\cite{Watanabe2017,Shiomi2019,Shiomi2020}. 
%just as the group velocity term of the anapole moment 
Hereafter, we call this mechanism of SCPE 'asymmetric origin' named after the asymmetric Bogoliubov spectrum.

%%%%alp_bet0 
\begin{figure*}[htbp]
  \centering
  \includegraphics[width=160mm]{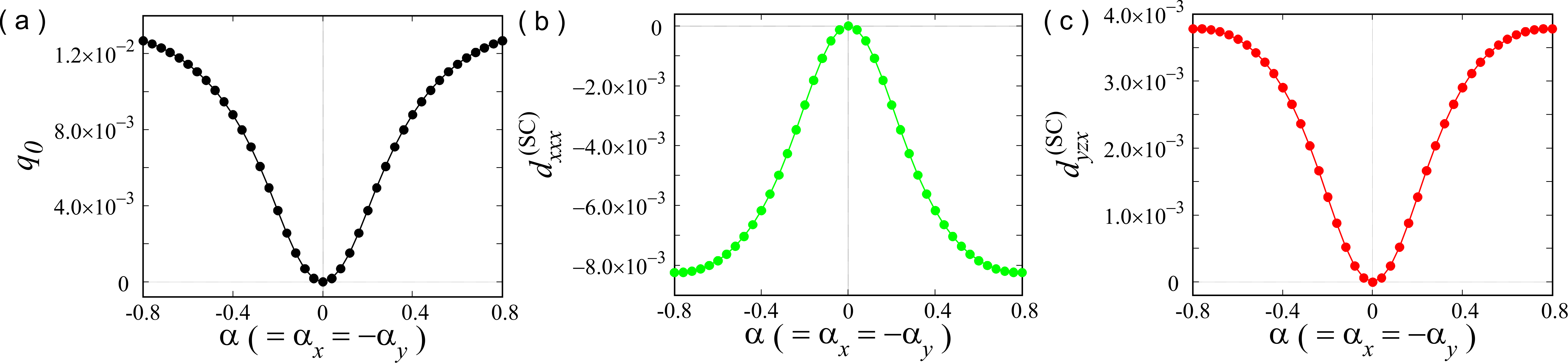}
  \caption{ 
 Results of the geometric effect model: $\alpha$ dependence of (a) $q_0$, (b) $d^{\textrm{(SC)}}_{xxx}$ in the $A_g + i B_{3u}$ anapole state, and (c) $d^{\textrm{(SC)}}_{yzx}$ in the $A_g + i A_{u}$ monopole state. We set $\alpha = \alpha_x = -\alpha_y$, %$\alpha_y = -\alpha_x = -\alpha$, 
 $\beta = 0$ and $T = 0.01$.
 }
\label{fig:alp_bet0}
\end{figure*}
%%%%tem_bet0
\begin{figure*}[tb]
  \centering
  \includegraphics[width=160mm]{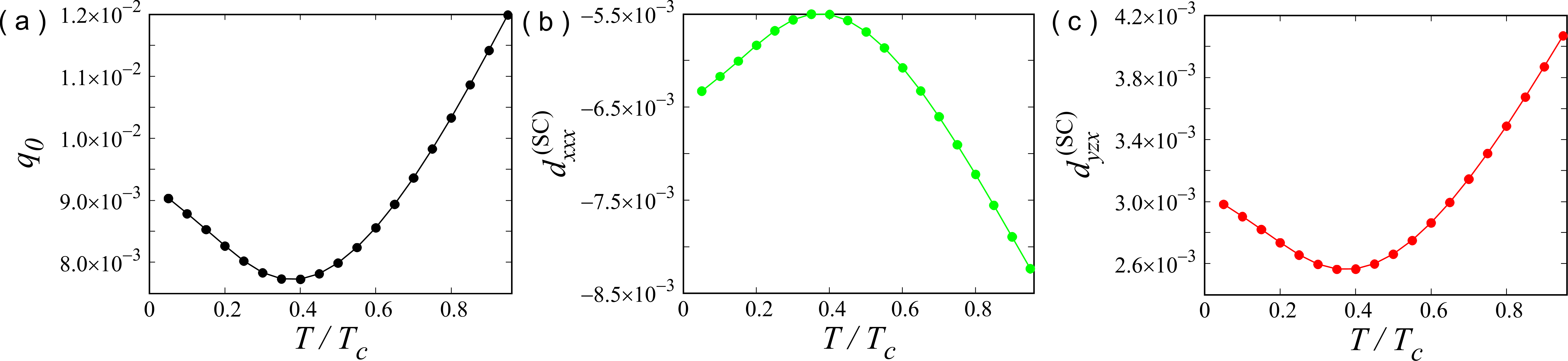}
  \caption{ 
 Results of the geometric effect model: temperature dependence of (a) $q_0$, (b) $d^{\textrm{(SC)}}_{xxx}$ in the $A_g + i B_{3u}$ anapole state, and (c) $d^{\textrm{(SC)}}_{yzx}$ in the $A_g + i A_{u}$ monopole state. We set $\alpha = \alpha_x = -\alpha_y = 0.4$ and $\beta = 0$.
 }
\label{fig:tem_bet0}
\end{figure*}

%%%%%%%% geometric model %%%%%%%%
\subsection*{ (2) Geometric effect model}
%%%%alp_bet0

Second, we set $\alpha_y = -\alpha_x (= -\alpha)$ and $\beta = 0$, in which the origin of the effective anapole moment is the geometric term since the group velocity term disappears.
Note that the point group of the monopole superconducting state with this parameter set is $D_4$, leading to the constraints $d^{\textrm{(SC)}}_{xyz} = 0$ and $d^{\textrm{(SC)}}_{yzx} = - d^{\textrm{(SC)}}_{zxy}$. 

The $\alpha$ dependence of $q_0$ and $d^{\textrm{(SC)}}_{ijk}$ is shown in Fig.~\ref{fig:alp_bet0}, %Both in the anapole and monopole states, 
which reveals the finite SCPE in the geometric effect model. However, in contrast to the group velocity model, the Bogoliubov spectrum is symmetric in this model for $q_0 =0$ (see Appendix~C). Indeed, $q_0 = 0$ in the monopole state and the spectrum is symmetric as $E_{\alpha}(\bm k) = E_{\alpha}(-\bm k) $.
%is realized in the stable monopole state, and the SCPE occurs.
Therefore, the origin of the SCPE must be different from the group velocity model where the asymmetric spectrum causes the SCPE.
%\NEW{Their $\alpha$-symmetric behavior in the anapole state is consistent with the anapole moment which is symmetric with respect to $\alpha$ in this model \cite{Kitamura2022}, as in the case of the group velocity model. On the other hand,}
%it is remarkable that the Bogoliubov spectrum is symmetric in this model for $q_0 =0$ (see Appendix~C). Especially, $q_0 = 0$ is realized in the stable monopole state, and the SCPE occurs. This indicates that the SCPE in this model is not due to the asymmetric Bogoliubov spectrum in contrast to the group velocity model discussed in the previous subsection. %and arises from the effect of quantum geometry. %from in the monopole state has nothing to do with the asymmetric spectrum. 
The geometric effect model also shows a similar parameter dependence of the Cooper pairs' momentum and the SCPE coefficients.
Considering the fact that the anapole moment arises from the quantum geometry in this model~\cite{Kitamura2022}, the similarity implies that the quantum geometry plays an essential role also in the SCPE. 
%などはどうでしょうか?
The $\alpha$ symmetric behavior of the SCPE coefficients $d^{\textrm{(SC)}}_{ijk}$ is consistent with this interpretation because the quantum geometry induces the $\alpha$ symmetric anapole moment in the geometric effect model model~\cite{Kitamura2022}.

The above discussion is supported by the temperature dependence of $q_0$ and $d^{\textrm{(SC)}}_{ijk}$ plotted in Fig.~\ref{fig:tem_bet0}.
Although there is a sizable energy gap in the spectrum (see Appendix~C), the SCPE and Cooper pairs' momentum are sizable even at low temperatures, in contrast to the group velocity model (Fig.~\ref{fig:tem_ay0}). 
Indeed, the temperature dependence of $q_0$ and $d^{\textrm{(SC)}}_{ijk}$ are weak.
This behavior is consistent with the above discussion because the effect of the quantum geometry of Bloch electrons is not suppressed by the energy gap. %leads to not only the anapole moment but also the SCPE. 
%These properties establish the existence of another origin of the SCPE 
Hereafter, we call this case of the SCPE 'symmetric origin' named after the symmetric Bogoliubov spectrum.

%%%%%%%% Mixed model %%%%%%%%
\subsection*{ (3) Mixed model}
%%%%alp_full
\begin{figure*}[htbp]
  \centering
  \includegraphics[width=160mm]{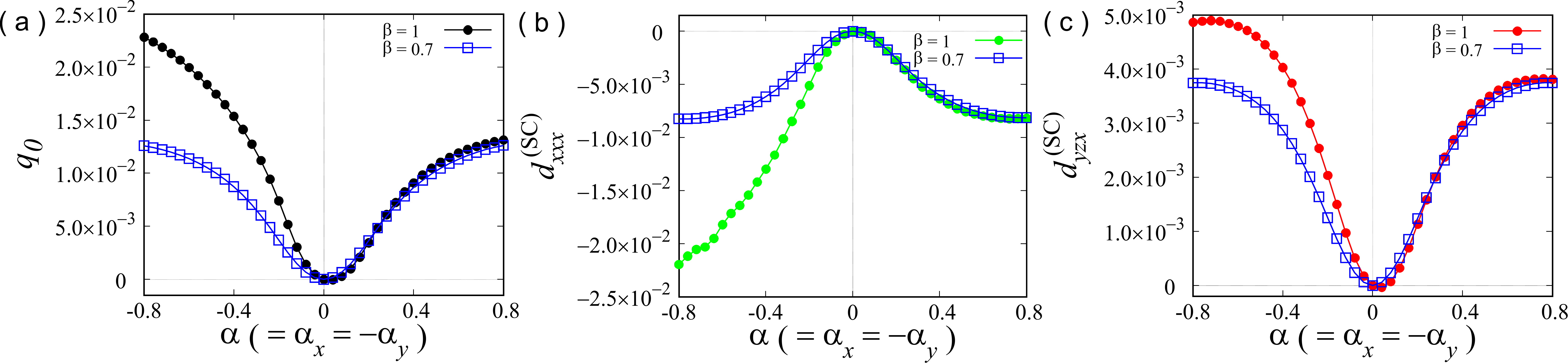}
  \caption{ 
 Results of the mixed model: $\alpha$ dependence of (a) $q_0$, (b) $d^{\textrm{(SC)}}_{xxx}$ in the $A_g + i B_{3u}$ anapole state, and (c) $d^{\textrm{(SC)}}_{yzx}$ in the $A_g + i A_{u}$ monopole state. We set $\alpha = \alpha_x = -\alpha_y$, %$\alpha_y = - \alpha_x = -\alpha$, 
 $\beta = 1$ or $0.7$, and $T = 0.01$. %The results with $\beta = 0.7$ are also shown for comparison.
 }
\label{fig:alp_full}
\end{figure*}

%%%%enedos_full
\begin{figure}[htbp]
  \centering
  \includegraphics[width=80mm]{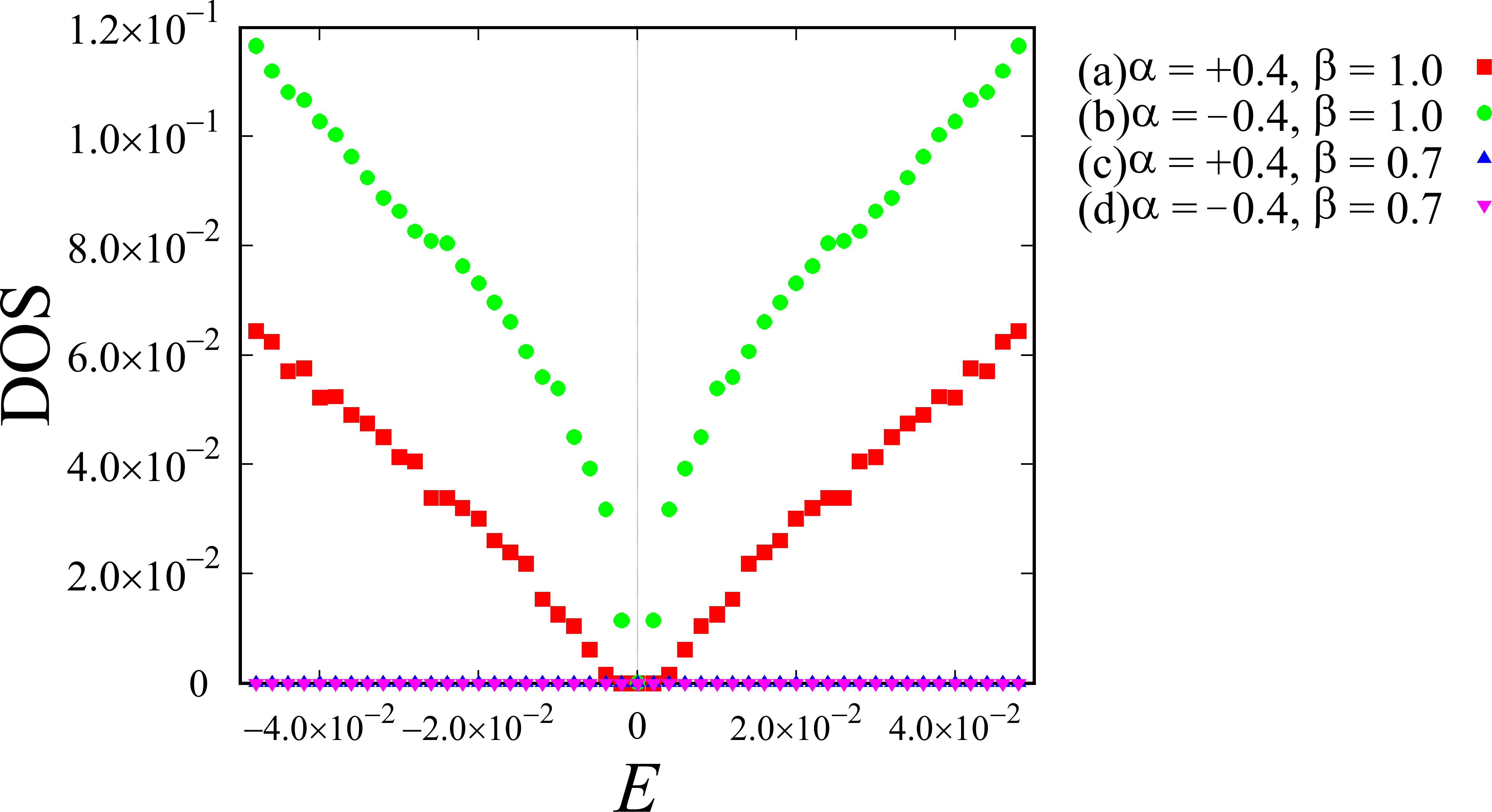}
  \caption{ 
The DOS in the $A_g + i B_{3u}$ anapole state with the following parameters: (a) $\alpha = +0.4$, $\beta = 1$, (b) $\alpha = -0.4$, $\beta = 1$, (c) $\alpha = +0.4$, $\beta = 0.7$, and (d) $\alpha = -0.4$, $\beta = 0.7$. We set $q_0 =0$ for simplicity.
 }
\label{fig:enedos_full}
\end{figure}
%%%%bet_full

%%%%bet_full
\begin{figure*}[htbp]
  \centering
  \includegraphics[width=160mm]{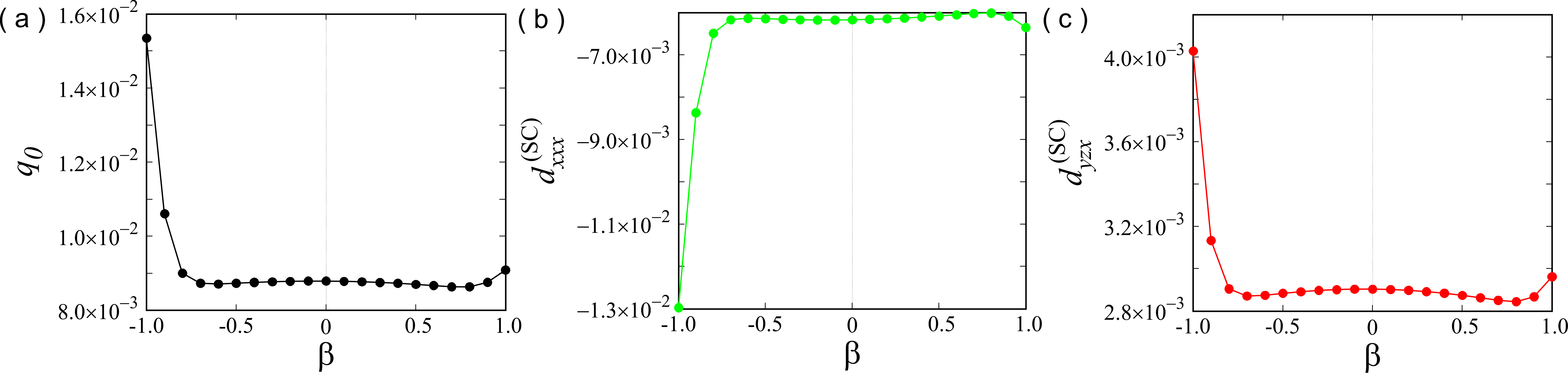}
  \caption{ 
 Results of the mixed model: $\beta$ dependence of (a) $q_0$, (b) $d^{\textrm{(SC)}}_{xxx}$ in the $A_g + i B_{3u}$ anapole state, and (c) $d^{\textrm{(SC)}}_{yzx}$ in the $A_g + i A_{u}$ monopole state. We set $\alpha = \alpha_x = - \alpha_y = 0.4$ and $T = 0.01$.
 }
\label{fig:bet_full}
\end{figure*}

Finally, we set $\alpha_y = -\alpha_x (=-\alpha)$ and $\beta \ne 0$, where both the group velocity term and the geometric term contribute to the effective anapole moment. As shown in Fig.~\ref{fig:alp_full}, the $\alpha$ dependence of $q_0$ and $d^{\textrm{(SC)}}_{ijk}$ is asymmetric for $\beta=1$ ,
%and this is consistent with the symmetry of the system. 
although it is almost symmetric for $\beta = 0.7$. We can interpret these features based on the results in the previous subsections: Since $q_0$ and  $d^{\textrm{(SC)}}_{ijk}$ are very small in the group velocity model for $\abs{\beta} < 1$ (Fig.~\ref{fig:bet_ay0}), the SCPE of the asymmetric origin is naturally small in the mixed model. Thus, the SCPE mainly arises from the symmetric origin, consistent with the $\alpha$-symmetric behavior similar to the geometric effect model (Fig.~\ref{fig:alp_bet0}). On the other hand, the asymmetric origin gives rise to a sizable contribution to the SCPE when $\abs{\beta} \geq 1$, making the SCPE asymmetric with respect to $\alpha$.

As we discussed for the group velocity model, the SCPE of the asymmetric origin is related to the DOS. This is correct in the mixed model as well. 
%This interpretation can be confirmed by calculating the DOS. As shown in 
Figure~\ref{fig:enedos_full} shows the DOS in the mixed model. First, we see that the superconducting gap suppresses the low-energy DOS for $\beta = 0.7$, consistent with the negligible contribution to the SCPE by the asymmetric origin. 
%there is few DOS with $\alpha = \pm 0.40, \beta = 0.70$ in this region. This is consistent with the small contribution of the asymmetric origin. In addition, while there is a large DOS with $\alpha = -0.40, \beta = 1.0$, there is few DOS with $\alpha = +0.40, \beta = 1.0$ in $\abs{E} < 0.004$. 
Second, we see sizable DOS for $\beta =1$, and it is larger for $\alpha=-0.4$ than for $\alpha=0.4$.
%This suggests the asymmetric origin with the later parameters is smaller than that with the former parameters. 
%This is also consistent with the fact that for $\alpha > 0$ the SCPE coefficients for $\beta =0.7$ almost coincide with those for $\beta=1$. 
Thus, it is indicated that the SCPE of the asymmetric origin is suppressed for $\alpha > 0$ because of the small low-energy DOS. In other words, we see a significant contribution of the asymmetric origin when $\beta \geq 1$ and $\alpha <0$. This is consistent with the parameter dependence of the SCPE in Fig.~\ref{fig:alp_full}. 
%results with the later parameters are almost the same as those with $\alpha = +0.40, \beta = 0.7$, while the former parameters give very different results from $\alpha = -0.40, \beta = 0.7$ as shown in Fig.~\ref{fig:alp_full}.

The $\beta$ dependence also supports the above discussion. 
Note that the parameter sets ($\alpha_x, \alpha_y, \beta$) and ($-\alpha_x, -\alpha_y, -\beta$) give the same results in our model.
%\YY{We can show that the SCPE is equivalent for sets of parameters $(\alpha,\beta)$ and $(-\alpha,-\beta)$.}
Fig.~\ref{fig:bet_full} shows the $\beta$ dependence of the SCPE coefficients for $\alpha =0.4$. The drastic change around $\beta=-1$ is attributed to the sizable DOS, which is equivalent to that for $(\alpha,\beta)=(-0.4,1)$. Figure~\ref{fig:bet_full} also reveals that the SCPE of the symmetric origin is nearly $\beta$-independent.
%As shown in Fig.~\ref{fig:bet_full}, they drastically change around $\beta = -1.0$ and this is consistent with the large contribution of the asymmetric origin with $\alpha = -0.40, \beta = 1.0$. On the other hand, they change only a little in the other region. This behavior reveals the weak $\beta$-dependence of the symmetric origin.

%\YYS{We note that this is not a simple summation of $q_0$ or the SCPE coefficients in the (1) group velocity model and (2) geometric effect model because the coexistence of $\alpha_y$ and $\beta$ change the contribution of asymmetric and symmetric origin in the models discussed above. However, our results strongly suggest that both $q_0$ and the SCPE coefficients can be divided into these two contributions, just as the effective anapole moment.} 

\begin{figure*}[htbp]
  \centering
  \includegraphics[width=160mm]{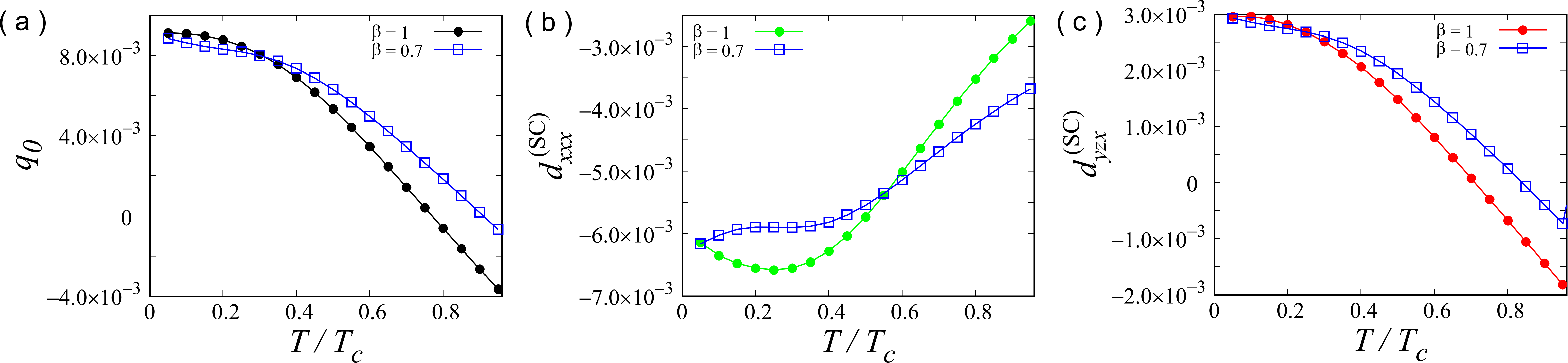}
  \caption{ 
 Results of the mixed model: temperature dependence of (a) $q_0$, (b) $d^{\textrm{(SC)}}_{xxx}$ in the $A_g + i B_{3u}$ anapole state, and (c) $d^{\textrm{(SC)}}_{yzx}$ in the $A_g + i A_{u}$ monopole state. We set $\alpha = \alpha_x = - \alpha_y = 0.4$ and $\beta = 1$ or $0.7$.
 }
\label{fig:tem_full}
\end{figure*}

The SCPE in the mixed model may show a unique temperature dependence as a consequence of the competition between the asymmetric and symmetric origins.  As shown in Figs.~\ref{fig:tem_full}(a) and \ref{fig:tem_full}(c), $q_0$ and $d^{\textrm{(SC)}}_{yzx}$ change the sign at a certain temperature. The sign reversal occurs because  the temperature dependence is significantly different between the SCPE of the asymmetric origin and that of the symmetric origin
%Since the $T$ dependence is different between the asymmetric and symmetric origin 
(compare Fig.~\ref{fig:tem_ay0} with Fig.~\ref{fig:tem_bet0}).
When the superconducting state is gapped, the SCPE is dominated by the symmetric origin at low temperatures, and thus, the quantum geometry is expected to play an essential role. On the other hand, the asymmetric origin related to the asymmetric Bogoliubov spectrum gives a large contribution near the transition temperature, and it can cause the sign change. 
%they cancel out each other if their sign is different and can be zero. 
Note that the sign reversal is not a general property, and it is sensitive to the detail of the system and SCPE mode. Indeed, there is no sign change in $d^{\textrm{(SC)}}_{xxx}$ (see Fig.~\ref{fig:tem_full}(b)), for instance.

It should be noticed that the Cooper pairs' momentum $q_0$ and the SCPE coefficients $d^{\textrm{(SC)}}_{ijk}$ show similar behaviors in all the models and parameters studied in this paper. This is also the case of the helical superconducting state studied earlier \cite{Chazono2022}. Surprisingly, this correspondence applies to the SCPE in the monopole superconducting state as well. Although $q_0=0$ in the monopole state, the SCPE shows a similar parameter dependence to $q_0$ in the anapole state. From these results, we suppose that the SCPE arises from the asymmetric Bogoliubov spectrum and quantum geometry like the anapole moment~\cite{Kitamura2022}. 
Further analysis of the SCPE related to quantum geometry is left as a future issue.
%To clarify the microscopic mechanism of the SCPE is key to understand this relation, but we remain this as a future issue. 

%\subsection{Notable other modes}
%Almost modes listed in Table~\ref{table:Mode} show qualitatively similar behavior with $d^{\textrm{(SC)}}_{xxx}$ in the $A_g + i B_{3u}$ state and $d^{\textrm{(SC)}}_{yzx}$ in the $A_g + i A_{u}$ state. However, there are some modes showing unique behavior. For example, $d^{\textrm{(SC)}}_{xyz}$ in the $A_g + i A_{u}$ state is vanished in the geometric effect model even if we set $\alpha_x \ne -  \alpha_y$ and the symmetry allows the realization of this mode (\textcolor{red}{Why?}). Reflecting this property, the mixed model is mainly characterized by the group velocity term as shown in Fig.~\ref{fig:dxyz}.

%\textcolor{red}{I want to mention about $d^{\textrm{(SC)}}_{xyy}$ and $d^{\textrm{(SC)}}_{zxy}$ modes which show $\beta$-linear dependence. BUT the interpretation is not established.}

%%%%%dxyz
%\begin{figure}[tb]
%  \centering
%  \includegraphics[width=70mm]{dxyz.pdf}
%  \caption{ 
% (a) $\alpha$ dependence and (b) $T$ dependence of $d^{\textrm{(SC)}}_{xyz}$ in $A_g + i A_{u}$ state in the (1) group velocity model.  (c) $\alpha$ dependence and (d) $T$ dependence in the (3) mixed model. We here set $\alpha_y = - \alpha$
% }
%\label{fig:dxyz}
%\end{figure}

%%%%%%%% Superconducting diode effect in the anapole superconductor %%%%%%%%
\section{Field-free diode effect}\label{sec:SDE}

%%%%sde
\begin{figure}[tb]
  \centering
  \includegraphics[width=65mm]{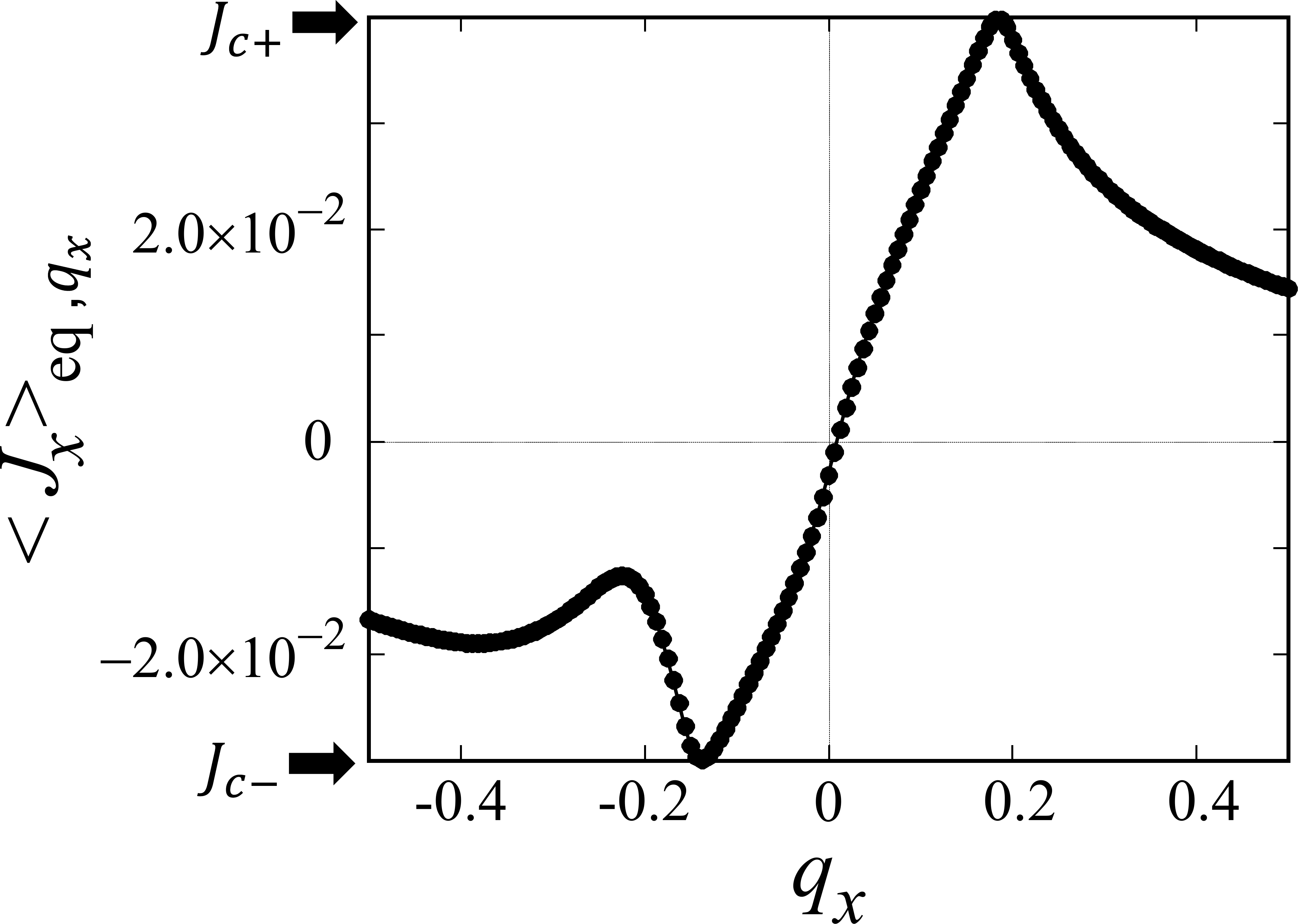}
  \caption{ 
 Supercurrent $\ave{\hat{J}_{x}}_{\textrm{eq}, q_x}$ as a function of $q_x$ in the mixed model of the $A_g + i B_{3u}$ anapole state. We set $\alpha = \alpha_x = - \alpha_y = 0.4$, $\beta = 1$, and $T = 0.01$. Critical currents $J_{c +}$ and $J_{c -}$ are marked by arrows.
 }
\label{fig:qx_jx}
\end{figure}

In this section, we demonstrate the field-free SDE in the anapole superconductors, which means the nonreciprocity in the critical current in the absence of the magnetic field. Here, we consider the SDE along the {\it x}-axis with the $A_g+iB_{3u}$ anapole superconducting state in mind. Adopting the formulation for the intrinsic SDE \cite{Daido2022}, we calculate the depairing critical current as
\begin{align}
J_{\textrm{c} +} = \textrm{max}_{q_x} \ave{\hat{J}_{x}}_{\textrm{eq}, q_x} ,~~ J_{\textrm{c} -} = \textrm{min}_{q_x} \ave{\hat{J}_{x}}_{\textrm{eq}, q_x},
\label{Jc}
\end{align}
using Eq.~\eqref{Jave}. The $q_x$ dependence of $\ave{\hat{J}_{x}}_{\textrm{eq}, q_x}$ in the $A_g+iB_{3u}$ anapole state is shown in Fig.~\ref{fig:qx_jx} for example, %\YYS{Because we give the superconducting gap phenomenologically instead of determining in the self-consistent treatment,  $\ave{\hat{J}_{x}}_{\textrm{eq}, q_x}$ remains finite with all $q_x$ region unlike the previous study \cite{Daido2022}.} 
by which we determine $J_{\textrm{c} +}$ and $J_{\textrm{c} -}$. %confirming the sufficiently wide $q_x$ region. 
The nonreciprocal component of the critical current is given by
\begin{align}
\Delta J_{\textrm{c}} = J_{\textrm{c} +}  + J_{\textrm{c} -},
\label{DJc}
\end{align}
and the SDE efficiency is defined as
\begin{align}
r = \Delta J_{\textrm{c}} / \bar{J_{\textrm{c}}},
\label{SCDE}
\end{align}
with $\bar{J_{\textrm{c}}} = (J_{\textrm{c} +}  - J_{\textrm{c} -}) / 2$. 

\begin{figure*}[tb]
  \centering
  \includegraphics[width=160mm]{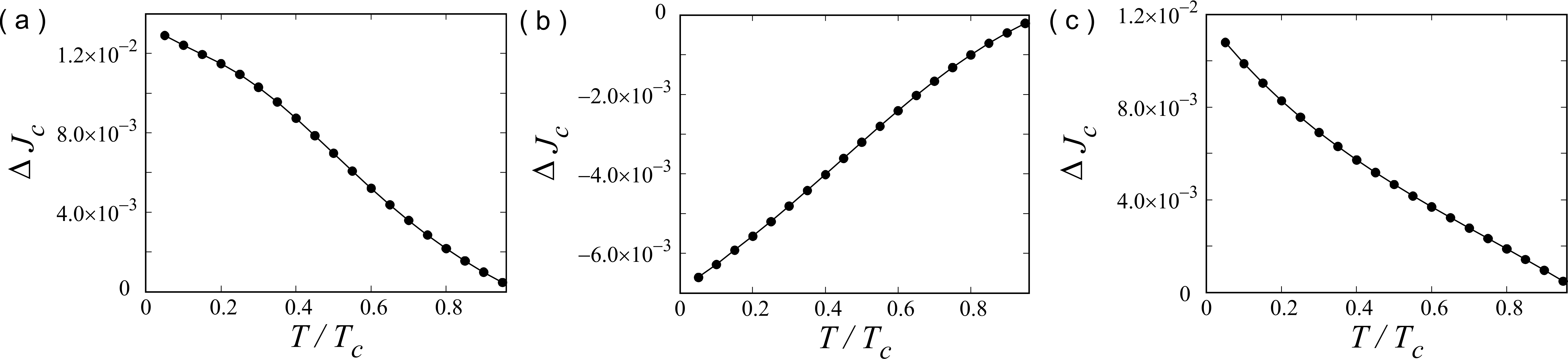}
  \caption{ 
 Temperature dependence of the nonreciprocal critical current $\Delta J_{\textrm{c}}$ in the (a) group velocity model, (b) geometric effect model, and  (c) mixed model of the $A_g + i B_{3u}$ anapole state. The parameters $(\alpha, \beta)$ are (a) $(0.4,1)$, (b) $(0.4,0)$, and (c) $(0.4,1)$
% the same 
as in Figs.~\ref{fig:tem_ay0}, \ref{fig:tem_bet0} and \ref{fig:tem_full}. %respectively.
 }
\label{fig:tem_sde}
\end{figure*}

\begin{figure*}[tb]
  \centering
  \includegraphics[width=160mm]{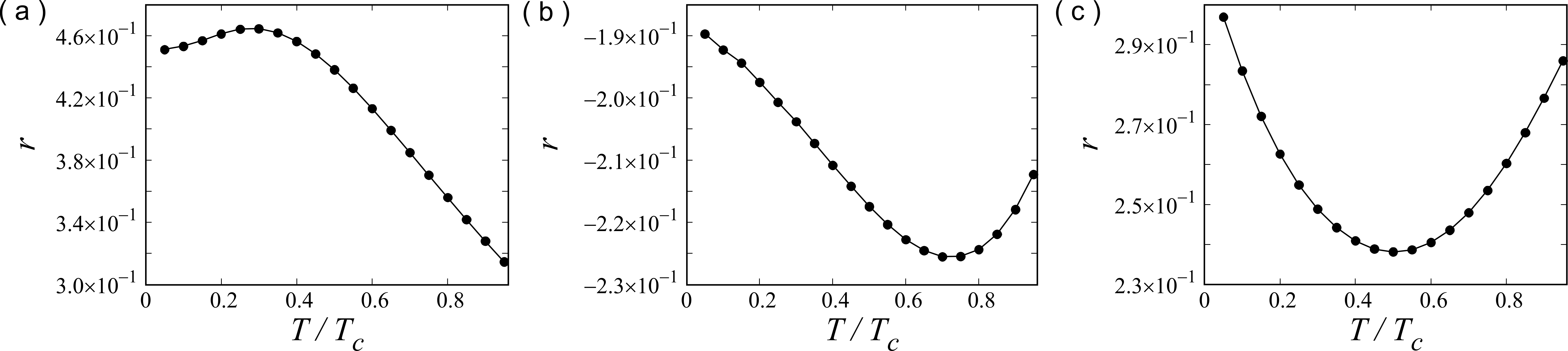}
  \caption{ 
 Temperature dependence of the SDE efficiency $r$ in the (a) group velocity model, (b) geometric effect model, and  (c) mixed model of the $A_g + i B_{3u}$ anapole state. The parameters $(\alpha, \beta)$ are (a) $(0.4,1)$, (b) $(0.4,0)$, and (c) $(0.4,1)$
as in Figs.~\ref{fig:tem_ay0}, \ref{fig:tem_bet0} and \ref{fig:tem_full}. 
 }
\label{fig:tem_nrr}
\end{figure*}

The numerical results of $\Delta J_{\textrm{c}}$ and $r$ are shown in Figs.~\ref{fig:tem_sde} and \ref{fig:tem_nrr}, respectively. We obtain finite nonreciprocity in the critical current characterized by $\Delta J_{\textrm{c}}$ and $r$ in all the models for the anapole superconducting states, namely, the (1) group velocity model, (2) geometric effect model, and (3) mixed model. %discussed in the previous section. 
Thus, the field-free SDE is a ubiquitous feature of anapole superconductors.
It is shown that $\Delta J_{\textrm{c}}$ is suppressed monotonically with increasing temperature. This behavior is in stark contrast to the fact that the temperature dependence of Cooper pairs' momentum $q_0$ and SCPE coefficients significantly depends on the model. %\YYS{while it is consistent with the previous study \cite{Daido2022}.} 
Note that the temperature scaling around $T=T_{\rm c}$ is not reliable because the $q$-dependence of the magnitude of gap function is neglected in our calculation, while it is negligible and the results are reliable at low temperatures \cite{Daido2022}.
Interestingly, the SDE efficiency $r$ reaches 40\%, which is comparable to the maximum value in the helical superconducting state at high magnetic fields \cite{Daido2022}. Thus, our results suggest a sizable SDE in the anapole superconducting state at the zero magnetic field.
%%%%tem_sde
%%%%tem_nrr

The SDE along the {\it x}-axis is allowed only in the anapole state, and it vanishes in the monopole state. Generally speaking, the SDE occurs in the anapole state with supercurrent in the same direction as the Cooper pairs' momentum $q_0$. Therefore, the SDE is suitable as a probe to distinguish the anapole and monopole states and to determine the direction of the anapole moment.

%%%%%%%% end Superconducting diode effect in the anapole superconductor %%%%%%%%

%%%%%%%% Summary and Discussion %%%%%%%%
\section{SUMMARY AND DISCUSSION}
\label{sec:Summary}
In this paper, we formulated and demonstrated the superconducting piezoelectric effect (SCPE) in the anapole and monopole superconducting states. 
We also showed the field-free superconducting diode effect (SDE) in the anapole superconducting state. 
The spontaneous IS and TRS breaking in these $PT$-symmetric superconducting states allows the off-diagonal and nonreciprocal responses without external symmetry-breaking fields. 
Therefore, the SCPE and SDE directly reflect the symmetry of superconducting states, and they can be used for probing the exotic symmetry and topology of superconductors.
In particular, the SCPE occurs under all the symmetry groups lacking the IS and TRS. Thus, in principle, we can distinguish the symmetry of superconducting states by the analysis of the SCPE tensor. For instance, we provided the classification table for the SCPE tensor in the $D_{2h}$ point group, corresponding to the candidate superconductor UTe$_2$. On the other hand, the field-free SDE occurs only along the anapole moment in the anapole superconducting state. Therefore, the observation of the SDE may evidence the anapole superconductivity and determine the direction of the anapole moment. 

Our calculations revealed the close relationship between the SCPE and the Cooper pairs' momentum. In the anapole state, the Cooper pairs can get finite momentum like in the FFLO and helical superconducting states, and the momentum is proportional to the anapole moment. In our results, the SCPE coefficients show similar parameter dependence to the Cooper pairs' momentum. 
According to our recent studies \cite{Kanasugi2022,Kitamura2022}, the anapole moment and Cooper pairs' momentum may have several origins, namely, the asymmetric spectrum of Bogoliubov quasiparticles and quantum geometry of Bloch electrons. The similarity implies the same origins of the SCPE. 
%of the effective anapole moment. In all cases, we obtained the finite SCPE coefficients and our results can be interpreted that the SCPE has two origins corresponding to the anapole moment \cite{Kitamura2022}, even in the monopole state. We also found a close relation between the SCPE and Cooper pairs' total momentum. 
Interestingly, the relation is confirmed between the SCPE in the monopole state and Cooper pairs' momentum in the anapole state. 
To clarify the microscopic origin of the SCPE, further theoretical analysis is desired and remains to be a future issue.
%To understand such unique properties, it is essential to clarify the microscopic origin of the SCPE. 

%We also revealed that the superconducting diode effect (SDE) can be realized in the anapole superconductor. Regardless of the origin of the anapole moment, the nonreciprocal ratio is relatively large compared to that in the helical superconducting state \cite{Daido2022}. 
On the other hand, the SDE looks unrelated to the Cooper pairs' momentum unlike the results of the helical superconducting state \cite{Daido2022,Daido2022-2}.
Therefore, the SDE is unlikely to be used for a probe of the magnitude of Cooper pairs' momentum. 
A characteristic property of the anapole superconducting state is that the SDE occurs at the zero magnetic field. 
Such field-free SDE has been searched in the recent research of SDE, but the platform is limited at present~\cite{Narita2022, Lin2021, Scammell2022, Wu2022}. Anapole superconductors are a platform of field-free SDE without symmetry-breaking magnetic order or external fields.
The rectified supercurrent is parallel to the momentum of Cooper pairs. 
In our calculation, a large SDE quality factor over 40\% is obtained.
%we cannot confirm the relation between the SDE and $q_0$. It is required to investigate the SDE in the anapole state with a self-consistent treatment to clarify whether the relation is essentially not valid or merely disappeared due to the simplification.

We expect that the SCPE and SDE will be complementary to other observable quantities characterizing the exotic superconducting stats. 
For instance, our recent work \cite{Kitamura2022} proposed a phenomenon specific to the anapole superconductor, the temperature-dependent Bogoliubov Fermi surface. The Bogoliubov Fermi surface affects thermodynamic properties~\cite{Autti2020, Setty2020, Ahn2021}, which could be experimentally verified, in principle. On the other hand, the SCPE and SDE occur regardless of the presence or absence of the Bogoliubov Fermi surface.

An intriguing future task is to examine the SCPE and SDE in UTe$_2$, a candidate of the anapole and monopole superconductivity. A recent ultrasound measurement detected softening of the elastic mode corresponding to the strain $s_{zx}$ \cite{Yanagisawa2022}. Thus, it is expected that the corresponding SCPE mode is enhanced. That is the $B_1$ mode in the anapole superconducting state while $B_2$ mode in the monopole superconducting state. The former is induced by the supercurrent along the {\it z}-axis, and it is along the {\it y}-axis in the latter.
%In addition, the SCPE can be observed even in the monopole state and the SDE distinguishes between the anapole state and monopole state. This study provides the SCPE and SDE as additional tools to investigate the anapole and monopole superconductors.

%%%%%%%% end Summary and Discussion %%%%%%%%

%%%%%%%% Acknowledgments %%%%%%%%
\begin{acknowledgments}
We thank A. Daido for fruitful discussion. This work was supported by JST SPRING (Grant Number JPMJSP2110), JSPS KAKENHI (Grants No.~JP18H01178, No.~JP18H05227, No.~JP20H05159, No.~JP21K18145, No.~JP22H01181, No.~JP22H04933, No.~JP22J22520) and SPIRITS 2020 of Kyoto University.
\end{acknowledgments}
%%%%%%%% end Acknowledgments %%%%%%%%

%%%%%%%% Appendix %%%%%%%%
\appendix
\section{Symmetry analysis of $PT$-symmetric superconducting states and SCPE based on the point group $D_{2h}$}
We discuss the anapole and monopole superconducting states classified based on the $D_{2h}$ point group. There are four even-parity and four odd-parity irreducible representations in the $D_{2h}$ point group, and accordingly, their coexistence allows sixteen mixed-parity superconducting states.
%ways of the parity mixing of the order parameters. 
The classification of superconducting states is summarized in Table~\ref{table:PT}. In the mixed-parity superconducting states, the point group symmetry is reduced from the normal state point group ($D_{2h}$) owing to the spontaneous parity violation. For instance, the point group of the $B_{1g} + i B_{3u}$ state is $C_{2v}$ with the principal axis in the $y$ direction. This means that the $B_{1g} + i B_{3u}$ state is an anapole superconducting state, where Cooper pairs can get total momentum in the $y$ direction, $\vq = (0, q_0, 0)$. In Table~\ref{table:PT}, we represent $C_{2v} (y)$ for such symmetry. The table reveals that the point group of parity-mixed superconducting states may be either $C_{2v} (x)$, $C_{2v} (y)$, $C_{2v} (z)$ (anapole), or $D_2$ (monopole). We have shown the SCPE mode in the $C_{2v} (x)$ anapole and $D_2$ monopole states in Tables~\ref{table:Ana} and \ref{table:Mono}, respectively. For completeness, we show the possible SCPE mode in the $C_{2v} (y)$ and $C_{2v} (z)$ anapole superconducting states in Tables~\ref{table:Anay} and \ref{table:Anaz}, respectively. 

%\begin{table}[t]
% \caption{Point group of the superconducting states realized when the parity mixing occurs in $D_{2h}$ systems. $x, y$ and $z$ denote the direction of the principal axis.}
% \label{table:AM}
% \centering
%  \begin{tabular}{c||c|c|c|c|}
%  \hline
%    & $A_g$ & $B_{1g}$ & $B_{2g}$ & $B_{3g}$ \\
%  \hline \hline
%   $A_u$ & $D_2$ & $C_{2v} (z)$ & $C_{2v} (y)$ & $C_{2v} (x)$ \\
%  \hline
%   $B_{1u}$ & $C_{2v} (z)$ & $D_2$ & $C_{2v} (x)$ & $C_{2v} (y)$ \\
%  \hline
%   $B_{2u}$ & $C_{2v} (y)$ & $C_{2v} (x)$ & $D_2$ & $C_{2v} (z)$\\
%  \hline
%   $B_{3u}$ & $C_{2v} (x)$ & $C_{2v} (y)$ & $C_{2v} (z)$ & $D_2$ \\
%  \hline
%  \end{tabular}
%\end{table}
\begin{table}[htbp]
 \caption{List of the IRs of the $C_{2v} (y)$ anapole superconducting state and corresponding strain $s_{ij}$, supercurrent $J_k$, and SCPE mode $d^{\textrm{(SC)}}_{ijk}$.}
 \label{table:Anay}
 \centering
  \begin{tabular}{ccc||c}
  \hline
   IR & Strain & Supercurrent & SCPE modes\\
  \hline \hline
   $A_{1}$ & $s_{xx}, s_{yy}, s_{zz}$ & $J_y$ & $d^{\textrm{(SC)}}_{xxy}~ d^{\textrm{(SC)}}_{yyy}~ d^{\textrm{(SC)}}_{zzy}$ \\
  \hline
   $A_{2}$ & $s_{zx}$ & $-$ & $-$ \\
  \hline
   $B_{1}$ & $s_{xy}$ & $J_x$ & $d^{\textrm{(SC)}}_{xyx}$ \\
  \hline
   $B_{2}$ & $s_{yz}$ & $J_z$ & $d^{\textrm{(SC)}}_{yzz}$ \\
  \hline
  \end{tabular}
\end{table}
\begin{table}[htbp]
 \caption{List of the IRs of the $C_{2v} (z)$ anapole superconducting state and corresponding strain $s_{ij}$, supercurrent $J_k$, and SCPE mode $d^{\textrm{(SC)}}_{ijk}$.}
 \label{table:Anaz}
 \centering
  \begin{tabular}{ccc||c}
  \hline
   IR & Strain & Supercurrent & SCPE modes\\
  \hline \hline
   $A_{1}$ & $s_{xx}, s_{yy}, s_{zz}$ & $J_z$ & $d^{\textrm{(SC)}}_{xxz}~ d^{\textrm{(SC)}}_{yyz}~ d^{\textrm{(SC)}}_{zzz}$ \\
  \hline
   $A_{2}$ & $s_{xy}$ & $-$ & $-$ \\
  \hline
   $B_{1}$ & $s_{yz}$ & $J_y$ & $d^{\textrm{(SC)}}_{yzy}$ \\
  \hline
   $B_{2}$ & $s_{zx}$ & $J_x$ & $d^{\textrm{(SC)}}_{zxx}$ \\
  \hline
  \end{tabular}
\end{table}

%Next, we discuss the properties of the SCPE coefficients $d^{\textrm{(SC)}}_{ijk}$.
We here comment on the derivation of the possible SCPE modes. Tables~\ref{table:Ana}, \ref{table:Mono}, \ref{table:Anay}, and \ref{table:Anaz} are obtained by considering the condition that the supercurrent and strain belong to the same irreducible representation.
An alternative way is to consider the compatible relation of irreducible representations.
Because the SCPE tensor  $d^{\textrm{(SC)}}_{ijk}$ becomes finite with the reduction of the symmetry of the system, $d^{\textrm{(SC)}}_{ijk}$ must belong to the totally symmetric representation in the superconducting state and not in the normal state. When we apply this condition to the $D_{2h}$ point group, we find that finite SCPE coefficients $d^{\textrm{(SC)}}_{ijk}$  belong to the $B_{3u}$ ($A_u$) irreducible representation in the $A_g + i B_{3u}$ anapole ($A_g + i A_{u}$ monopole) state. %\YY{Tables~\ref{table:Ana}, \ref{table:Mono}, \ref{table:Anay}, and \ref{table:Anaz} are consistent with this criterion.}

%\YYS{Here, we discuss the parameter dependence of the SCPE coefficients based on these irreducible representations of $d^{\textrm{(SC)}}_{ijk}$.
%In the group velocity model, reversing the sign of $\alpha$ is equivalent to the exchange of sublattices and reversing the $x$-direction ($M_x$ mirror). %(the former changes the sign of $\alpha$ and $\beta$ in our model \eqref{BdGHamiltonianM}). 
%Since exchanging sublattices does not change the macroscopic physical property, we have only to consider the effect of the $M_x$ mirror operation. Because the $B_{3u}$ and $A_u$ irreducible representations are odd with this operation, the SCPE coefficients $d^{\textrm{(SC)}}_{ijk}$ are antisymmetric with respect to $\alpha$. We can understand the $\beta$-antisymmetric behavior in the same way (changing the sign of $\beta$ is equivalent to the reversal in the $x$-direction). 
%In contrast to the group velocity model, since changing the sign of $\alpha$ merely corresponds to the replacement of sublattice in the geometric effect model, it affects no results and $d^{\textrm{(SC)}}_{ijk}$ is $\alpha$-symmetric.}

%\YY{Is the above discussion correct?}

%\YYS{We cannot rewrite the sign change of $\alpha$ ($\beta$) as the combination of simple operations in the mixed model. This property leads to $\alpha$- ($\beta$-) asymmetric behavior.}

\section{Analysis of the group velocity model}
\begin{figure}[htbp]
  \centering \includegraphics[width=80mm]{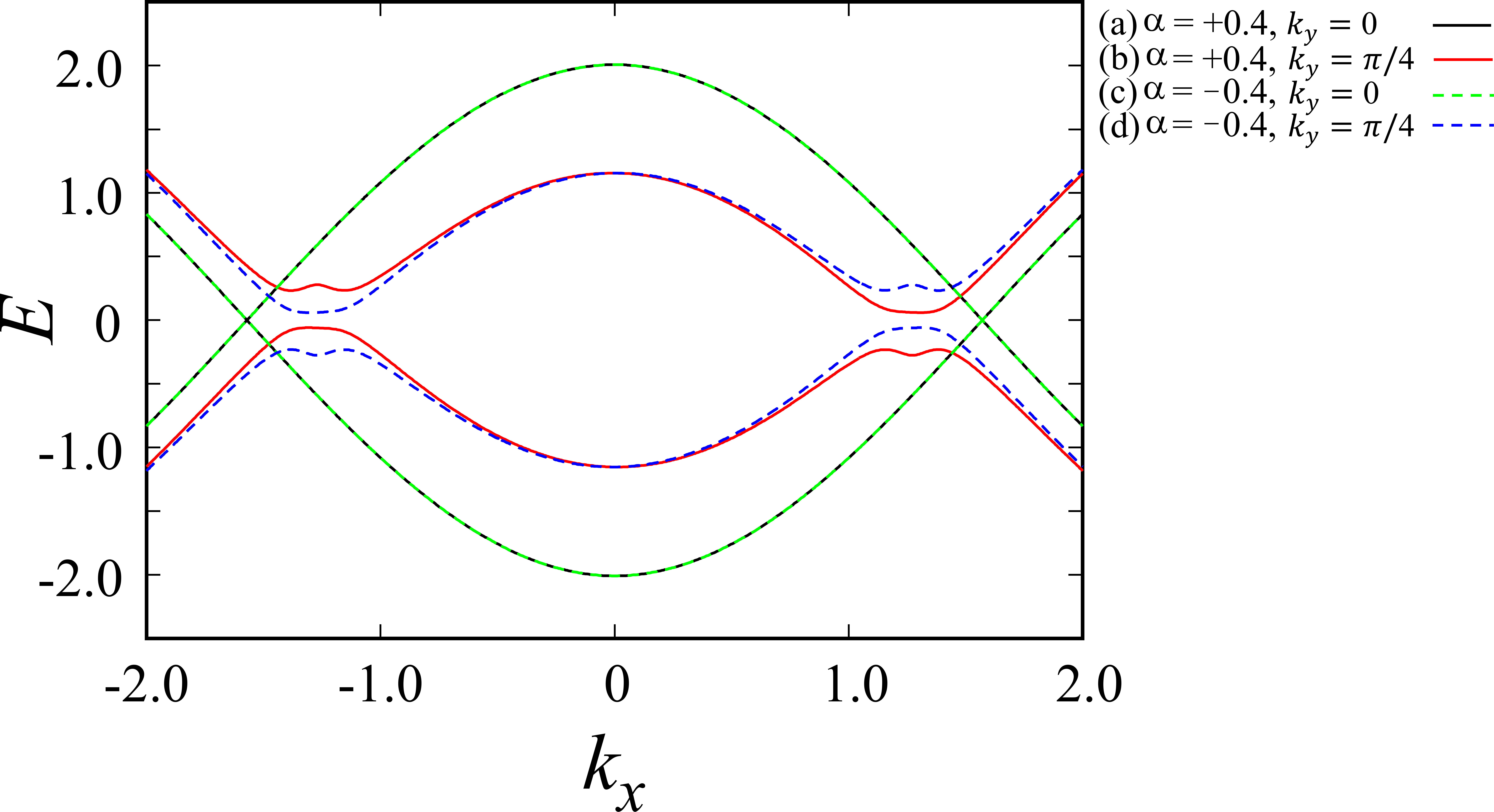}
  \caption{ 
The spectrum of Bogoliubov quasiparticles in the group velocity model for the $A_g + i B_{3u}$ anapole state.
We show the energy dispersion along the $k_x$ axis by setting $k_z = 0$ and (a),(c) $k_y = 0$ or (b),(d) $k_y = \pi/4$. 
The spectrum with $k_y = 0$ is gapless.
(a),(b) The solid lines represent the spectrum calculated with $\alpha_x = 0.4$, while (c),(d) the dashed lines are obtained with 
$\alpha_x = -0.4$. We assume $\alpha_y = 0$, $\beta = 1$, $T=0.01$ and set $q_0 = 0$. 
 }
\label{fig:ABS_ay0}
\end{figure}

In this Appendix, we show some notable properties of the group velocity model. First, we show the Bogoliubov spectrum in Fig.~\ref{fig:ABS_ay0} assuming Cooper pairs' momentum $q_0=0$. The spectrum is asymmetric in the $k_x$ direction and the asymmetry is reversed by changing the sign of the ASOC, $\alpha_x$ (see the spectrum for $k_y=\pi/4$). This property results in the $\alpha$-antisymmetric behavior of the SCPE and $q_0$ in the group velocity model. Their $\beta$-antisymmetric behavior is also explained in the same way.

%%%ABS_ay0

%%%BFS_ay0
\begin{figure}[htbp]
  \centering
\includegraphics[width=65mm]{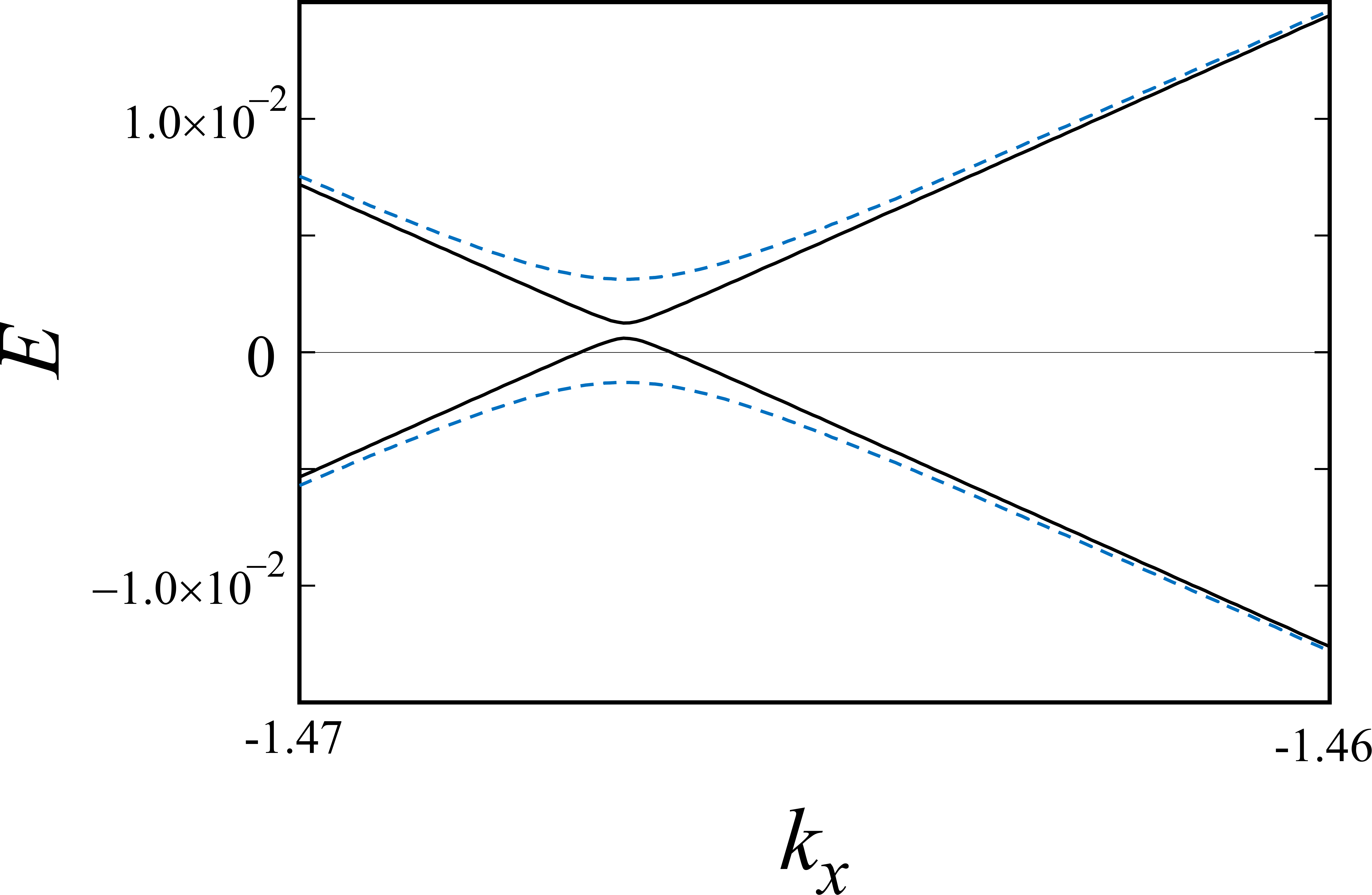}
  \caption{ 
%The spectrum of Bogoliubov quasiparticles in the group velocity model for the $A_g + i B_{3u}$ anapole state. 
The same plot as Fig.~\ref{fig:ABS_ay0}.
The black solid (blue dashed) line shows the result for $\alpha_x = 0.4$ and $\beta = 1.01$ ($\beta = 1$). We set $k_y = -0.0481$ and $k_z = -0.457$ to show a part of the Bogoliubov Fermi surface. 
 }
\label{fig:BFS_ay0}
\end{figure}

Next, we discuss the gap structure.
Note that the spectrum is symmetric on the $k_y = 0$ and $k_y = \pi$ planes if we set $q_0 = 0$ (see Fig.~\ref{fig:ABS_ay0} for $k_y=0$). The analytic representation of the Bogoliubov spectrum on these planes is obtained as
%of the energy spectrum by setting $q_0 = 0$. Since the spectrum at general  $\vk$ points is represented in a complicated form, we here focus on the $k_x$-$k_z$ plane with $k_y = 0$. Then, we obtain the energy of Bogoliubov quasiparticles
\begin{align}
E (\vk) = \pm \sqrt{(\varepsilon_{\vk} -\mu)^2 + \left(\abs{\Delta^{+}_{\vk}} \pm \beta \Delta_1 d^{g}_y\right)^2},
\label{E_velo}
\end{align}
where $\Delta^+_{\vk} = \Delta_1 \psi^g + \Delta_2 d^u_z$. There are nodes in the superconducting gap when $\varepsilon_{\vk} -\mu = 0$ and $\abs{\Delta^{+}_{\vk}} \pm \beta \Delta_1 d^{g}_y = 0$ are simultaneously satisfied on the planes. Indeed, for $\beta=1$, the point node is present on the $k_y = 0$ plane, as we see in Fig.~\ref{fig:ABS_ay0}.
When $\beta > 1$, Bogoliubov Fermi surfaces appear in several regions of the Brillouin zone. For instance, Fig.~\ref{fig:BFS_ay0} shows the Bogoliubov spectrum indicating the Bogoliubov Fermi surface for $\beta = 1.01$.
Thus, the superconducting gap structure significantly changes around $\beta=\pm 1$. This is the reason why the SCPE and $q_0$ show remarkable $\beta$ dependence around $\beta=\pm 1$ in the group velocity model.

\section{Bogoliubov spectrum in the geometric effect model}
\begin{figure}[htbp]
  \centering
  \includegraphics[width=65mm]{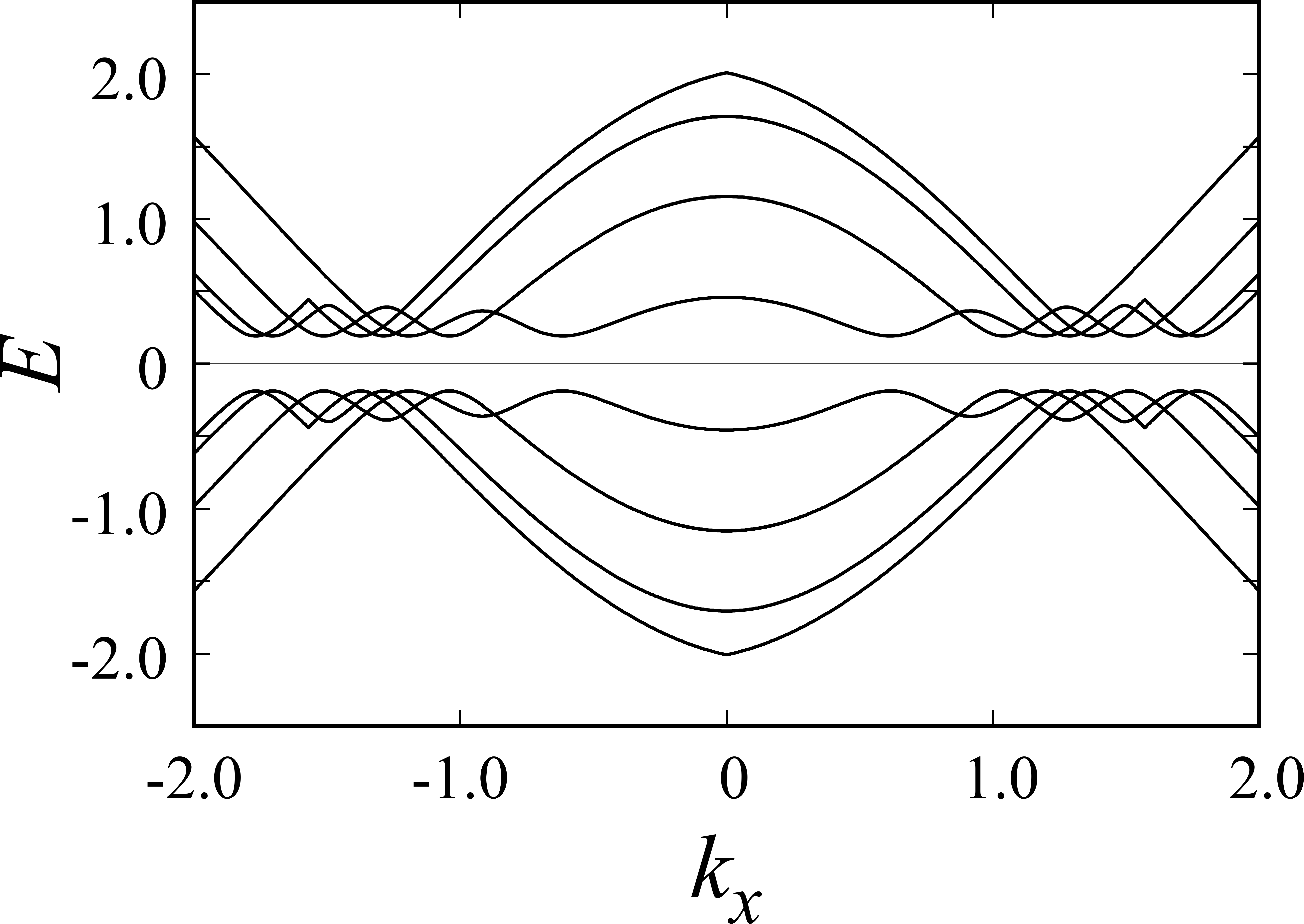}
  \caption{ 
The Bogoliubov spectrum of the $A_g + i B_{3u}$ state with $\alpha_x = - \alpha_y = 0.4$, $\beta = 0$ (geometric effect model), and $T = 0.01$. We draw the spectrum on $k_z = 0$ and $k_y = 0, \pi/8, \pi/4, 3\pi/8$.
 }
\label{fig:ABS_bet0}
\end{figure}

Here, we show the energy spectrum of Bogoliubov quasiparticles in the geometric effect model. The spectrum without $q_0$ can be analytically calculated and obtained as
\begin{align}
E (\vk) = \pm \sqrt{ \left( \abs{g^{+}_{\vk}} \pm \sqrt{(\varepsilon_{\vk} -\mu)^2 + (\textrm{Im} \Delta^{+}_{\vk})^2} \right)^2 + \left(\textrm{Re} \Delta^{+}_{\vk}\right)^2 },
\label{E_geo}
\end{align}
where $g^+_{\vk} = g_x + i g_y$. We confirm that the spectrum is symmetric for $\vk$, i.e. $E (\vk) = E (-\vk)$ in contrast to the group velocity model. In addition, since $\textrm{Re} \Delta^{+}_{\vk} = \Delta_1 \psi^g > 0$ is always finite in our model, gap nodes are absent at least for $q_0=0$, Indeed, we see the gapped spectrum in Fig.~\ref{fig:ABS_bet0}.

%%%ABS_bet0

\bibliography{article}

\end{document}